\documentclass[manuscript,screen]{acmart}

\usepackage[acronym]{glossaries}

\newacronym{dl}{DL}{Deep Learning}
\newacronym{amc}{AMC}{Automatic Modulation Classification}
\newacronym{dnn}{DNN}{deep neural network}
\newacronym{nn}{NN}{ Neural Network}
\newacronym{bn}{BN}{ Batch Normalization}
\newacronym{cnn}{CNN}{Convolutional Neural Network}
\newacronym{gap}{GAP}{Global Average Pooling}
\newacronym{fc}{FC}{Fully Connected}
\newacronym{snr}{SNR}{signal-to-noise ratio}
\newacronym{qos}{QOS}{Quality of Service}
\newacronym{5g}{5G}{Fifth Generation}
\newacronym{tl}{TL}{Transfer Learning}
\newacronym{fls}{FSL}{Few Shot Learning}
\newacronym{dg}{DG}{domain generalization}
\newacronym{tta}{TTA}{Test Time Adaptation}
\newacronym{iid}{IID}{independent and identically distributed}
\newacronym{uda}{UDA}{Unsupervised Domain Adaptation}
\newacronym{mse}{MSE}{mean squared error}

\newacronym{uav}{UAV}{unmanned autonomous vehicle}
\newacronym{xr}{XR}{extended reality}
\newacronym{vr}{VR}{virtual reality}
\newacronym{ar}{AR}{augmented reality}
\newacronym{cv}{CV}{computer vision}

\newacronym{ft}{FT}{Fine Tuning}
\newacronym{sfda}{SFDA}{Source-Free Domain Adaptation}
\newacronym{mac}{MAC}{multiply and accumulate}
\newacronym{nlp}{NLP}{natural language processing}

\newacronym{fgsm}{FGSM}{Fast Gradient Sign Method}
\newacronym{pgd}{PGD}{Projected Gradient Descent}
\newacronym{cw}{CW}{Carlini-Wagner}
\newacronym{lid}{LID}{Local Intrinsic Dimensionality}
\newacronym{jsma}{JSMA}{Jacobian-based Saliency Map Attack}
\newacronym{dp-means}{DP-means}{Dirichlet Process-Means}
\newacronym{acdc}{ACDC}{Automatic Circuit DisCovery}
\newacronym{nnif}{NNIF}{Nearest Neighbors Influence Functions}
\newacronym{ead}{EAD}{Elastic-net Attack to DNNs}
\newacronym{hop}{HOP}{Hopskipjump}
\newacronym{mmd}{MMD}{Maximum Mean Discrepancy}
\newacronym{hamper}{HAMPER}{Halfspace Mass Depth Detector}
\newacronym{ctt}{CTT}{Certifiable Taboo Trap}
\newacronym{eps}{EPS}{Expected Perturbation Score}
\newacronym{ihsd}{IHSD}{Intrinsic Hidden State Distribution}
\newacronym{ami}{AmI}{Attack Meets Interpretability}
\newacronym{frs}{FRS}{Face Reconginition Systems}
\newacronym{nic}{NIC}{Neural-network Invariant Checking}
\newacronym{bnn}{BNN}{Bayesian Neural Network}
\newacronym{libre}{LiBRe}{Lightweight Bayesian Refinement}
\newacronym{bu}{BU}{Bayesian Uncertainty}
\newacronym{ood}{OOD}{Out-of-Distribution}
\newacronym{id}{ID}{In-Distribution}
\newacronym{gda}{GDA}{Gaussian Discriminant Analysis}
\newacronym{mlloo}{ML-LOO}{ML-Leave-One-Out}
\newacronym{asv}{ASV}{Automatic Speaker Verification}
\newacronym{mfcc}{MFCC}{Mel-Frequency Cepstral Coefficients}
\newacronym{lcr}{LCR}{label change rate}
\newacronym{vit}{ViT}{Vision Transformer}
\newacronym{auroc}{AUROC}{Area Under Receiver Operating Characteristics}
\newacronym{kd}{KD}{Kernel Density}
\newacronym{kde}{KDE}{Kernel Density Estimator}
\newacronym{shap}{SHAP}{Shapely Additive Explanations}

\newacronym{odin}{ODIN}{ODIN}
\newacronym{godin}{GODIN}{Generalized ODIN}
\newacronym{dml}{DML}{Decoupling MaxLogit}
\newacronym{msp}{MSP}{Maximum Softmax Probability}
\newacronym{ln}{LogitNorm}{Logit Normalization}
\newacronym{react}{ReAct}{Rectified Activation}
\newacronym{gradnorm}{GradNorm}{Gradient Norm}
\newacronym{ashb}{ASH-B}{Activation Shaping by Binarizing}
\newacronym{mood}{MOOD}{Masked image modeling for OOD Detection}
\newacronym{oe}{OE}{Outlier Exposure}
\newacronym{doe}{DOE}{Distributional-agnostic Outlier Exposure}
\newacronym{divoe}{DivOE}{Diversified Outlier Exposure}
\newacronym{vra}{VRA}{Variational Rectified Activation}
\newacronym{nac}{NAC}{Neuron Activation Coverage}
\newacronym{ddim}{DDIM}{Denoising Diffusion Implicit Model}

\newacronym{ai}{AI}{artificial intelligence}
\newacronym{tpr}{TPR}{True Positive Rate}
\newacronym{fpr}{FPR}{False Positive Rate}
\newacronym{vmf}{vMF}{von Misses-Fisher}
\newacronym{in}{IN}{Instance Normalization}
\newacronym{mim}{MIM}{Momentum Iterative Method}
\newacronym{eot}{EOT}{Expectation Over Transformation}
\newacronym{aa}{AA}{Auto Attack}
\newacronym{bim}{BIM}{Basic Iterative Method}
\newacronym{gama}{GAMA}{Guided Adversarial Margin Attack}
\newacronym{fab}{FAB}{Fast Adaptive Boundary}
\newacronym{uap}{UAP}{Universal Adversarial Perturbation}
\newacronym{acg}{ACG}{Auto Conjugate Gradient}
\newacronym{lafeat}{LAFEAT}{Latent Feature Attack}
\newacronym{mt}{MT}{Multi-Target}
\newacronym{apgd}{APGD}{AutoPGD}
\newacronym{dlr}{DLR}{Difference of Logit Ratio}
\newacronym{nag}{NAG}{Network for Adversary Generation}
\newacronym{usgd}{USGD}{Universal Stochastic Gradient Descent}
\newacronym{cduap}{CD-UAP}{Class Descriminative UAP}
\newacronym{dfuap}{DF-UAP}{Dominant Feature UAP}
\newacronym{cuap}{C-UAP}{Cosine UAP}
\newacronym{sga}{SGA}{Stochastic Gradient Aggregation}
\newacronym{atuap}{AT-UAP}{Adversarially Trained UAP}
\newacronym{genap}{Gen AP}{Generative Adversarial Perturbation}
\newacronym{sgd}{SGD}{Stochastic Gradient Descent}
\newacronym{fff}{FFF}{Fast Feature Fool}
\newacronym{cwa}{CWA}{Common Weakness Attack}
\newacronym{tsea}{T-SEA}{Transfer-based Self-Ensemble Attack}
\newacronym{svre}{SVRE}{Stochastic Variance Reduced Ensemble}
\newacronym{me}{ME}{Model Ensemble}
\newacronym{gne}{GNE}{Ghost Network Ensemble}
\newacronym{rap}{RAP}{Reverse Adversarial Perturbation}
\newacronym{lbp}{LinBP}{Linear Back-Propagation}
\newacronym{nim}{NIM}{Nesterov Iterative Method}
\newacronym{dim}{DIM}{Diversed Input Method}
\newacronym{tim}{TIM}{Translation Invariant Method}
\newacronym{vt}{VT}{Variance Tuning}
\newacronym{sim}{SIM}{Scale Invariant Method}
\newacronym{gra}{GRA}{Gradient Relevant Attack}
\newacronym{taig}{TAIG}{Trasferable Attack based on Intergreted Gradient}
\newacronym{pgn}{PGN}{Penalizing Gradient Norm}
\newacronym{ssa}{SSA}{Spectrum Sanity Attack}
\newacronym{pam}{PAM}{Path-Augmented Method}
\newacronym{sit}{SIT}{Structure Invariant Transformation}
\newacronym{kegn}{KEGN}{Knowledge Extraction Generative Network}
\newacronym{gan}{GAN}{Generative Adversarial Network}
\newacronym{dast}{DaST}{Data-free Substitute Training}
\newacronym{dfme}{DFME}{Data-Free Model Extraction}
\newacronym{nes}{NES}{Natural Evolution Strategy}
\newacronym{zoo}{ZOO}{Zeroth-Order Optimization}
\newacronym{ba}{BA}{Boundary Attack}
\newacronym{sfa}{SFA}{Sign Flip Attack}
\newacronym{qeba}{QEBA}{Query-Efficient BA}
\newacronym{dct}{DCT}{Discrete Cosine Transform}
\newacronym{moo}{MOO}{Multi-Objective Optimization}
\newacronym{aca}{ACA}{Adversarial Content Attack}
\newacronym{ncf}{NCF}{Natural Color Fool}
\newacronym{sae}{SAE}{Sementic Adversarial Example}
\newacronym{wae}{WAE}{Wasserstein Adversarial Example}

\newacronym{at}{AT}{Adversarial Training}
\newacronym{kl}{KL}{Kullback-Leibler}
\newacronym{alp}{ALP}{Adversarial Logit Pairing}
\newacronym{cat}{CAT}{Curriculum Adversarial Training}
\newacronym{mma}{MMA}{Max-Margin Adversary}
\newacronym{yopo}{YOPO}{You Only Propagate Once}
\newacronym{vae}{VAE}{Variational Auto-Encoder}
\newacronym{milp}{MILP}{Mixed Integer Linear Programming}
\newacronym{smt}{SMT}{Satisfiability Modulo Theory}
\newacronym{lp}{LP}{Linear Programming}
\newacronym{sdp}{SDP}{Semi-Definite Programming}
\newacronym{ibp}{IBP}{Interval Bound Propagation}
\newacronym{awp}{AWP}{Adversarial Weight Perturbation}
\newacronym{da}{DA}{Domain Adaptation}
\newacronym{relu}{ReLU}{Rectified Linear Unit}
\newacronym{rbf-svm}{RBF-SVM}{Radial Basis Function Support Vector Machine}
\newacronym{ta}{TA}{Transferable attack}
\newacronym{svma}{SVMA}{Support Vector Machine Attack}
\newacronym{dta}{DTA}{Decision Tree Attack}
\newacronym{cdf}{CDF}{Cumulative Distance Function}
\newacronym{vi}{VI}{Value invariant}
\newacronym{pi}{PI}{Provenance Invariant}
\newacronym{knn}{k-NN}{k-nearest neighbor}
\newacronym{pca}{PCA}{Principal Component Analysis}
\newacronym{rle}{RLE}{Random-Local-Ensemble}
\newacronym{svm}{SVM}{Support Vector Machine}
\newacronym{rbf}{RBF}{Radial Basis Funtion}
\newacronym{sa}{SA}{Square Attack}
\newacronym{sta}{STA}{Spatial Transformation Attack}
\newacronym{lavan}{LaVAN}{Localized and Visible Adversarial Noise}
\usepackage{wrapfig}
\usepackage{fixltx2e}
\usepackage{multirow, booktabs}
\usepackage{pifont}

\usepackage{subfig}
\usepackage[skip=0pt]{caption}
\usepackage{lscape}
\usepackage{longtable}

\AtBeginDocument{%
  \providecommand\BibTeX{{%
    \normalfont B\kern-0.5em{\scshape i\kern-0.25em b}\kern-0.8em\TeX}}}

\setcopyright{acmlicensed}
\copyrightyear{2024}
\acmYear{2024}
\acmDOI{XXXXXXX.XXXXXXX}

\acmJournal{CSUR}

\begin{document}

\title[Resilience and Security of Deep Neural Networks Against Intentional and Unintentional Perturbations]{Resilience and Security of Deep Neural Networks Against Intentional and Unintentional Perturbations: Survey and Research Challenges}

\author{Sazzad Sayyed}
\authornotemark[3]
\authornote{These authors contributed equally to this research.}
\email{sayyed.a@northeastern.edu}
\author{Milin Zhang}
\authornotemark[1]\authornotemark[3]
\email{zhang.mil@northeastern.edu}
\author{Shahriar Rifat}
\authornotemark[1]\authornotemark[3]
\email{rifat.s@northeastern.edu}
\author{Ananthram Swami}
\authornotemark[2]
\email{ananthram.swami.civ@army.mil}
\author{Michael De Lucia}
\authornotemark[2]
\email{michael.j.delucia2.civ@army.mil}
\author{Francesco Restuccia}
\email{f.restuccia@northeastern.edu}\authornotemark[3]

\affiliation{%
  \institution{\newline $\ddagger$ Institute for the Wireless Internet of Things, Northeastern University}
  \city{Boston}
  \state{Massachussetts}
  \country{United States}
  \postcode{02120}
}
\affiliation{%
  \institution{\newline $\dagger$ DEVCOM Army Research Laboratory}
  \country{United States}
  \postcode{02120}
}

\renewcommand{\shortauthors}{Restuccia et al.}

\begin{abstract}
  In order to deploy \glspl{dnn} in high-stakes scenarios, it is imperative that \glspl{dnn} provide inference robust to external perturbations -- both intentional  and unintentional. Although the resilience of \glspl{dnn} to intentional and unintentional perturbations has been widely investigated, a unified vision of these inherently intertwined problem domains is still missing. In this work, we fill this gap by providing a survey of the state of the art and highlighting the similarities of the proposed approaches. We also analyze the research challenges that need to be addressed to deploy resilient and secure \glspl{dnn}. As there has not been any such survey connecting the resilience of \glspl{dnn} to intentional and unintentional perturbations, we believe this work can help advance the frontier in both domains by enabling the exchange of ideas between the two communities.      
\end{abstract}

\maketitle

\section{Introduction}

Thanks to their ability of performing critical tasks such as object detection \cite{wsrc2023}, language translation \cite{angelova-etal-2022-using}, image classification \cite{brock2021highperformance}, and efficient pose estimation \cite{maji2022yolopose}, \glspl{dnn} have become essential in our everyday life \cite{BOULEMTAFES2021100221}. For a comprehensive survey on \gls{dnn}s, we refer the reader to \cite{pouyanfar2018}. Breakthroughs in the form of large language models such as \textit{GPT-4} \cite{OpenAI_GPT4_2023}, universal segmentation models such as \textit{Segment Anything} \cite{kirillov2023segany} and diffusion models such as \textit{Stable Diffusion} \cite{Rombach_2022_CVPR} have   advanced the frontier of \gls{ai} and captured the interest of ordinary citizens in using \gls{ai} in their day-to-day lives. 

The unprecedented benefits of \glspl{dnn} come with their own set of social and ethical challenges, mainly in the form of privacy, safety and security issues \cite{elliott2022ai}. For example, it has been shown that a \gls{dnn} is extremely sensitive to intentional perturbations where changing a few pixels in the input can lead to misclassifications \cite{goodfellow2014explaining,moosavi2016deepfool,su2019one,Barrett_2020}. In addition, \glspl{dnn} are vulnerable to unintentional perturbations due to natural phenomena, for example, frost, rain, shot noise, impulse noise, defocus blur, motion blur and zoom blur, as well as natural shifts in distribution of labels, which is also known as \textit{semantic shift}. To ensure \gls{ai} systems based on \glspl{dnn} can be deployed in real-world systems, it is imperative to ensure resilience and security from \textit{both} intentional \textit{and} unintentional perturbations. Achieving this goal can be ensured, for example, by guaranteeing that the output of a \gls{dnn} will be inherently robust and accurate, or by implementing a rejection scheme that detects inputs on which the prediction is likely to be incorrect due to perturbation or semantic shift, so that the \gls{dnn} can be adapted to provide a correct prediction for such inputs under specific constraints. This makes the study of resilience and security of \glspl{dnn} a timely yet extremely complex issue. \vspace{-0.3cm}


\subsection{Motivation and Novel Contributions}

Although intentional and unintentional perturbations share several critical aspects, a unified vision  has so far been elusive. Since its inception in 2014 \cite{szegedy2014intriguing}, the study of resilient \gls{dnn} design has been fragmented into separate domains. Some approaches have been studying intentional perturbations -- also known as adversarial machine learning -- while some approaches have been proposed to guarantee robustness against \gls{ood} samples. Since a \gls{dnn} needs to be  resilient to  \textit{both} types of perturbation, it is important to study the connection between these domains. \cite{salehi2022unified} discusses the relation between \gls{ood} detection and anomaly detection, open-set recognition, and novel set recognition. The survey in \cite{aldahdooh2022adversarial} focuses on the detection of adversarial inputs and bench-marking some of the detection approaches. To the best of our knowledge, literature lacks surveys focusing on the connection between the detection of adversarial input and \gls{ood} input. As such, in this paper we  discuss the literature from the perspective of resilience of \glspl{dnn} encompassing detection of both adversarial and \gls{ood}. Specifically, the main contributions of this paper can be summarized as follows:

\begin{itemize}
\vspace{-0.1cm}
    \item We categorize and discuss the seminal, significant, and recent work in the domain of \gls{ood} detection (i.e., unintentional interference) and adversarial sample detection (i.e., intentional interference);
    \item We investigate the commonalities among intentional and unintentional perturbation detection and the corresponding defense strategies, while remarking the strengths and weaknesses of these approaches. We believe these two communities can benefit from this study as they can discover common approaches, similarities in these two fields, and adopt new perspective from the other community.
    \item We conclude the paper by pointing out some open questions and research directions regarding ensuring the resilience of \glspl{dnn} in real-world inference systems. \vspace{-0.2cm}
\end{itemize}

\section{Vulnerabilities of Deep Neural Networks: Background and Taxonomy}


We define the term "vulnerability"  as any action that causes the \gls{dnn} to perform in not its intended manner as compared to when the action is absent. Under the scope of this survey, we consider  actions where external perturbations are introduced to the input samples during inference. In this section, we provide background on different approaches to introduce these perturbations intentionally. Next, we illustrate different scenarios that occur when external perturbations are added to the input samples in a natural fashion. \vspace{-0.3cm}      

\subsection{Intentional Perturbation: Adversarial Attack}

\begin{figure}
    \centering
    \includegraphics[width=0.95\textwidth]{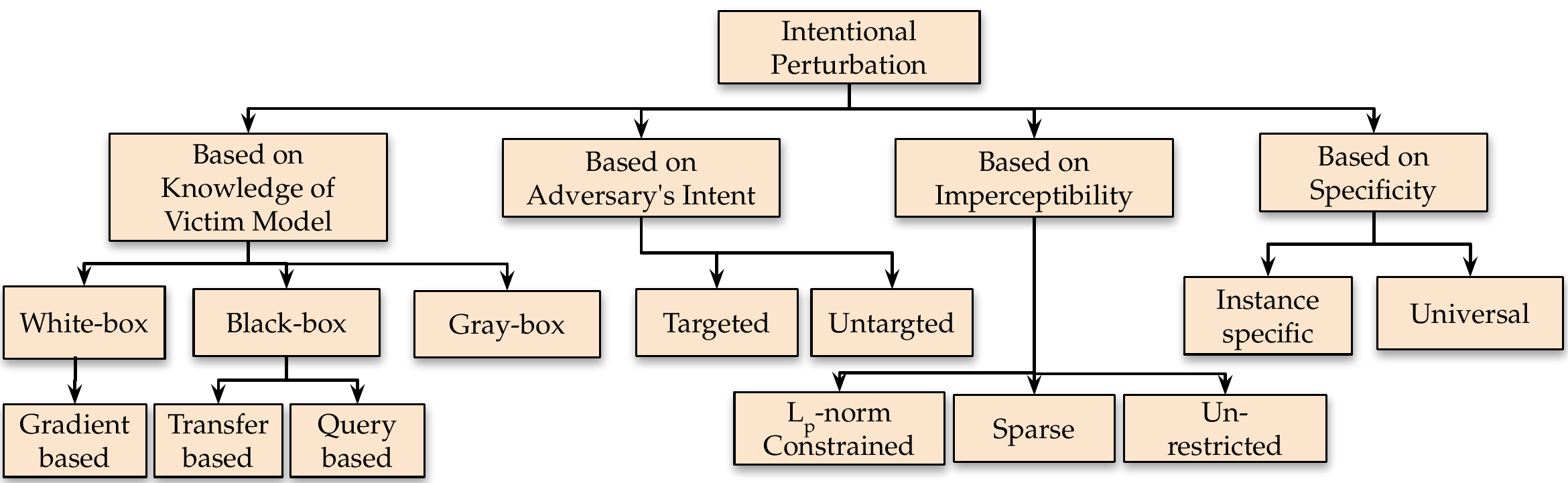} \vspace{0.3cm}
    \caption{Taxonomy of intentional perturbations. \vspace{-1cm}}
    \label{fig:atk}
\end{figure}

We consider a \textit{target \gls{dnn}} $f(x)=y$, where $x\in \mathcal{X}$ and $y\in \mathcal{Y}$ represent input and output samples respectively, with $\mathcal{X}$, $\mathcal{Y}$ respectively representing input and output space. The objective of an adversary is to find a perturbation $\delta$ that misleads the target DNN. Based on their intent or objectives, adversarial attacks can be categorized as \textbf{untargeted} and \textbf{targeted}. The former aim at causing an incorrect classification without specifying a particular target class, i.e., 
\begin{equation}
    f(x+\delta) \neq y, \quad \delta \leq \epsilon . \label{eqn:untarget_atk}
\end{equation}
Equation \ref{eqn:untarget_atk} describes the untargeted attack scenario where $\epsilon$ is a constraint determining the magnitude of the perturbation. On the other hand, the targeted attack aims at inducing a specific incorrect classification toward a specific class $\hat{y}\neq y$, 
\begin{equation}
    f(x+\delta) = \hat{y}, \quad \delta \leq \epsilon . \label{eqn:target_atk}
\end{equation}
Based on the attacker's knowledge of the target \gls{dnn}, attacks can be classified as \textbf{white-box} \cite{goodfellow2014explaining,kurakin2018adversarial,carlini2017towards,madry2018towards,athalye2018synthesizing}, \textbf{black-box} \cite{papernot2016transferability,papernot2017practical,liu2017delving,ilyas2018black,andriushchenko2020square}, and \textbf{gray-box} \cite{wang2021gray-box, lapid2023see}. In the white-box setting, the attacker has full knowledge of the targeted \gls{dnn} architecture, parameters, and training data, which enables highly-effective perturbations. A common strategy in white-box attacks involves formulating the attack as an optimization problem, facilitating the utilization of gradient descent to identify adversarial examples. Consequently, these techniques are often referred to as \textbf{gradient-based} attacks. For example, \gls{fgsm} \cite{goodfellow2014explaining} uses a single step in the direction of gradient of the loss function $\mathcal{L(\cdot)}$ with respect to the input to obtain the necessary perturbation to evade correct detection of the the input sample as shown in equation \ref{eqn:fgsm}.~\gls{bim}  \cite{kurakin2018adversarial} improved the approach by searching iteratively. Its variant \gls{pgd} \cite{madry2018towards} with random initialization and multiple restarts is considered as a baseline approach to assess adversarial robustness in literature. 
\begin{equation}
    x_{adv} = x + \epsilon \cdot sgn(\triangledown_x \mathcal{L}(\theta,x,y)) , \label{eqn:fgsm}
\end{equation}

\gls{cw} \cite{carlini2017towards} demonstrated the importance of the loss function and optimizer. Furthermore, they proposed a surrogate loss given by Equation \ref{eqn:marginloss},
\begin{equation}
    \mathcal{L}(\mathbf{z},y) = \underset{z_i\neq y}{\max}~z_i - z_y. \label{eqn:marginloss}
\end{equation}
which  denotes the margin loss where $\mathbf{z}$ is the output vector, $z_i$ is the score of $i$-th class of the target \gls{dnn} and $y$ is the ground-truth label of the sample. \gls{mt} \cite{gowal2019alternative} enhances the gradient attack by a surrogate loss function designed for multiple targeted classes $\tau \in T$,
\begin{equation}
    \mathcal{L}(\mathbf{z},y) = \underset{\tau \in T}{\sum}z_\tau - z_y, \quad y \notin T. \label{eqn:mtloss}
\end{equation}
Equation~\ref{eqn:mtloss} denotes the multi-targeted loss, which aims at optimizing adversarial examples across multiple targeted classes and thus bypass the local minima associated with a single class. As a result, it demonstrates strong capability to evade defense algorithms. Among other notable work \gls{eot} \cite{athalye2018synthesizing} employs input augmentation, \gls{gama} \cite{sriramanan2020guided} utilizes guidance from function mapping of unperturbed samples, \gls{aa} \cite{croce2020reliable} uses ensemble of attacks, \gls{acg} \cite{yamamura2022diversified} applies conjugate directions to guide the gradient descent, and MIFPE \cite{yu2023efficient} minimizes effect of floating point error on the gradient step.

On the other hand, black-box attacks have only limited knowledge about the victim model. As a result, black-box attacks leverage indirect information such as model transferability and output score to generate effective adversarial examples. Black-box attacks encompass a diverse array of strategies, including \textbf{transfer-based}~\cite{xie2019improving,Lin2020Nesterov,dong2019evading}, and \textbf{query-based}~\cite{chen2017zoo,ilyas2018black,andriushchenko2020square} attacks, among others. In the transfer attack scenario, attackers perform gradient attacks on a group of surrogate models by assuming that adversarial samples misleading one \gls{dnn} model are also likely to mislead others. Therefore, transfer attacks consists of two parts: surrogate model training and transferable adversarial sample generation. \gls{me} \cite{liu2017delving} first proposed to use averaged prediction from ensemble of models to improve transferability. \gls{cwa} \cite{chen2024rethinking} targets the common weaknesses of multiple surrogate models based on cosine similarity and smoothness. Other notable approaches are self-ensemble (\gls{gne} \cite{li2020learning}, \gls{tsea} \cite{huang2023t}), input transformations (\cite{xie2019improving}, \cite{Lin2020Nesterov}, Admix \cite{wang2021admix}, \gls{ssa} \cite{long2022frequency}, \gls{sit} \cite{wang2023structure}, \gls{pam} \cite{zhang2023improving}), gradient calibration (\gls{mim} \cite{dong2018boosting}, \gls{nim} \cite{Lin2020Nesterov}, \gls{lbp} \cite{guo2020backpropagating}, \gls{pgn} \cite{ge2024boosting}, \gls{tim} \cite{dong2019evading}, \gls{taig} \cite{huang2022transferable}, \gls{vt} \cite{wang2021enhancing}, \gls{taig} \cite{huang2022transferable},\gls{gra} \cite{zhu2023boosting}), and surrogate model training (\gls{kegn} \cite{yoo2019knowledge}, \gls{dfme} \cite{zhang2021data}, \gls{dast} \cite{zhou2020dast}).  

In query-based attacks, attackers can only access the input and corresponding \gls{dnn} output (\textit{i.e.}, probability scores or hard labels). Adversarial samples are generated based on an iterative search by sending multiple quires to the target model. The approaches of query based attacks follow one of two streams - \textit{score based} and \textit{decision based} scenario. \cite{chen2017zoo} first proposes zeroth order optimization approach and applies gradient estimation to find perturbations without substitute model training. \gls{nes} \cite{ilyas2018black}, N-attack \cite{li2019nattack}, AdvFlow \cite{mohaghegh2020advflow}, and NP-attack \cite{bai2020improving} try to approximate the gradient information. As gradient estimation demands a large number of queries, SimBA \cite{guo2019simple}, Square Attack \cite{andriushchenko2020square}, PPBA \cite{li2020projection}, and BABIES\cite{tran2022exploiting} utilize random search to find the optimal perturbation for attacking the model. 

Gray-box adversarial attacks, also known as partial knowledge attacks, are a class of adversarial machine learning techniques where the attacker has limited knowledge about the target model. Unlike black-box attacks, where the attacker has no knowledge of the model's architecture or parameters, and white-box attacks, where the attacker has full access to the model, gray-box attacks assume the attacker has some information—such as knowledge of the model's architecture but not its exact parameters, or access to some but not all training data.  \cite{papernot2017practical} demonstrated that adversarial examples crafted for a substitute model (one that approximates the target model) can effectively transfer to the target model, indicating the potency of gray-box attacks . Another study by \cite{tramèr2018ensemble} explored ensemble methods to enhance the robustness of models against gray-box attacks by using multiple substitute models to generate adversarial examples, showing that such attacks can still circumvent defenses designed for black-box or white-box scenarios .

Based on the imperceptibility of perturbations, adversarial attacks can be classified as \textbf{$L_p$ norm constrained}~\cite{goodfellow2014explaining,kurakin2018adversarial,madry2018towards}, \textbf{sparse}~\cite{croce2019sparse,modas2019sparsefool,dong2020greedyfool}, and \textbf{unrestricted}~\cite{xiao2018spatially,Bhattad2020Unrestricted,shamsabadi2020colorfool} attacks. In most of the literature, the threat model is bounded in $l_2$ or $l_\infty$ norm. In other word, the constraint in Eqn.~\ref{eqn:untarget_atk} and Eqn.~\ref{eqn:target_atk} are refined as $||\delta||_p \leq \epsilon$, where $p=2$ or $p=\infty$. Sparse attacks investigate a more difficult problem where attackers can only perturb a small number of pixel of each input. Unrestricted attacks leverage small digital transformations of inputs (\textit{i.e.}, rotation, scaling, brightness) or generate noise that contains semantic difference in physical world (\textit{i.e.}, shadows, raindrops, laser beams) to attack victim models. 

Based on the specificity of perturbations, adversarial attacks can be classified as \textbf{instance-aware}~\cite{goodfellow2014explaining,kurakin2018adversarial,madry2018towards} and \textbf{universal} attack~\cite{moosavi2017universal,mopuri2018nag,li2019stealthy, zhang2020cd}. Instance-aware attack aims to generate adversarial examples tailored to each specific instance. This approach considers the unique characteristics of each input when crafting perturbations to maximize their effectiveness. Universal adversarial perturbation, on the other hand, is a perturbation that can be applied universally to a wide range of samples, regardless of their specific characteristics. Unlike instance-aware perturbations, universal perturbations are designed to have a broad impact across different instances, potentially affecting various types of inputs. They are crafted to exploit vulnerabilities in machine learning models consistently across diverse datasets or input distributions.

\subsection{Unintentional Perturbation: Determining Out-of-Distribution Samples}

\begin{figure}[!h]
    \centering
    \includegraphics[width=0.95\textwidth]{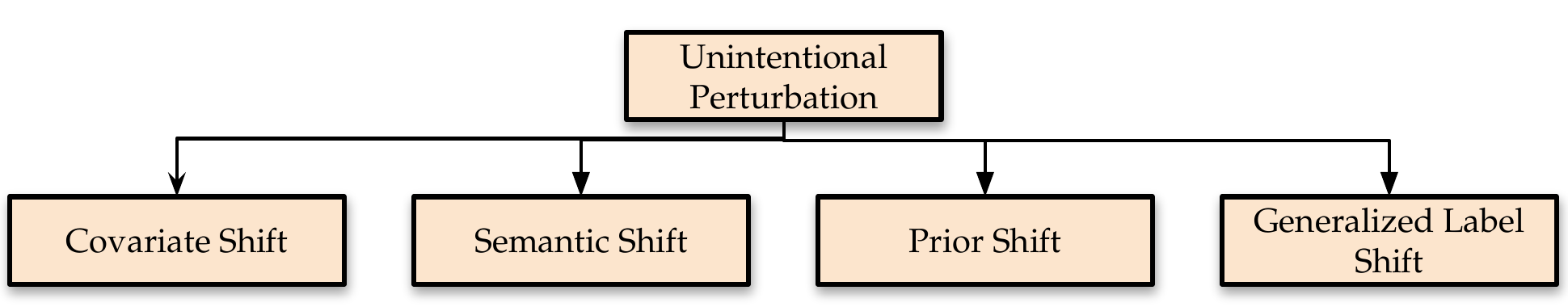}
    \caption{Taxonomy of unintentional perturbation. \vspace{-0.2cm}}
    \label{fig:perturbation}
\end{figure}

Figure \ref{fig:perturbation} shows a taxonomy of existing work in unintentional perturbation. To study the resilience of \gls{dnn} to unintentional perturbation, existing work makes assumption regarding the nature of the perturbation. If the perturbation has no resemblance to the distribution of the training data, it is extremely challenging to adapt the \gls{dnn} without additional knowledge and/or labeled samples. In this case, perturbation detection can be considered as a viable option.

 We introduce some notation and define $\mathit{X}$ and $\mathit{Y}$ as the random variables respectively representing input and labels. $\mathcal{D}_S(X)$ and $\mathcal{D}_S(Y)$ is used to denote the marginal distribution of input samples and the labels in the domain from where the training dataset is sampled.  The joint distribution of input and label $\mathcal{D}_S(X, Y)$ is consequently known as \textit{training distribution}, \textit{source distribution} or \textit{in-domain distribution}. Similarly, $\mathcal{D}_T(X)$ and $\mathcal{D}_T(Y)$ are used to denote the marginal distribution of input and label in the target domain while their joint distribution $\mathcal{D}_T(X,Y)$ is termed as the \textit{target distribution}, or \textit{out-of-distribution}. We define the input space $\mathcal{X}\in \mathbb{R}^d$ and the corresponding label space $\mathcal{Y}$, where $d$ indicates the dimension of the input. The dimension of $\mathcal{Y}$ varies depending on the task, e.g., for classification tasks, $\mathcal{Y}=[C]\triangleq \{1, ...,C\}$, where C denotes the number of classes, and $\mathcal{Y}\in \mathbb{R}^d$ for semantic segmentation tasks. We define a \gls{dnn} $f_{\theta_S}:\mathcal{X} \mapsto\mathcal{Y}$ composed of a feature extractor $g_{\theta_S}:\mathcal{X} \mapsto\mathcal{Z}$ and a predictor $h_{\theta_S}:\mathcal{Z} \mapsto\mathcal{Y}$ such that $f=h\circ g$, where $\mathcal{Z}$ is the intermediate representation space and $\mathit{Z}$ is the corresponding random variable. 

Out of distribution (OOD) samples can be defined as any input drawn from any distribution different from training distribution. The term \textit{distribution shift} encompasses four types of shifts in the literature as summarized in Figure \ref{fig:perturbation}. The first is \textit{non-semantic} or \textit{covariate shift} (e.g. noise, blur, frost and different types of corruptions, change in geographical location, brightness etc.) and the second is \textit{semantic shift} or \textit{label set shift}, which corresponds to the emergence of new semantic category or class label.  The third is \textit{prior shift}, and corresponds to a shift in the distribution of the classes observed.  For example, an object detection algorithm will be more likely to encounter ``animal'' objects in a rural area than in an urban area. The fourth is the \textit{generalized label shift}, where the prior shift and the covariate shift happens simultaneously. We formally define the four types of distribution shifts below.

\noindent \textbf{Definition of Covariate Shift}: The \textit{covariate shift} is a change in the distribution $\mathcal{D}(X)$ which preserves their label distribution $\mathcal{D}(Y)$. In other words, $\mathcal{D}_S(Y|X) = \mathcal{D}_T(Y|X)$. Notice that the target feature distribution $\mathcal{D}_T(X)$ is assumed to be known. Since the covariate shift does not change the semantic content of the input, it is also named \textit{non-semantic shift}. Direct perturbation, either intentional or unintentional, may cause such shift. Among the unintentional perturbations, we have common corruptions such as Gaussian blur, Gaussian noise, motion blur, defocus blur, frost, fog and rain, among others \cite{hendrycks2019robustness}. This also includes shift in input distribution due to a change in geographical location, operating hardware (e.g., different camera or sensors) or viewpoint \cite{hendrycks2021many}. Due to the change in the distribution $\mathcal{D}(\mathcal{X})$, the distribution $\mathcal{D}(\mathcal{X},\mathcal{Y})$ changes. This brings the problem of dataset shift \cite{quinonero2008dataset, peng2019moment}. Additionally, we can consider intentionally perturbed inputs a.k.a. adversarial inputs \cite{moosavi2016deepfool, carlini2017towards, madry2018towards} as covariate shifts, since the perturbation only changes the input distribution. This highlights the connection between unintentional and intentional perturbations. 

\noindent \textbf{Definition of Semantic Shift:} Any kind of shift that changes the semantic content and as a result the marginal distribution of label $\mathcal{D}(\mathcal{Y})$ of the dataset is a semantic shift or label shift. A semantic shift affects both the image space and the label space as the distribution of input $\mathcal{D}_{X}$ is shifted from the source distribution and new labels are introduced in the label space. The detection of such shifts can be considered as encompassing the tasks of \textit{novel class recognition} and \textit{open set recognition}. In the {open set recognition} problem \cite{vaze2022openset}, the \gls{dnn} is presented with sample from classes which were not present in training data representing a shift in the label distribution. One-class {novel class recognition} can be thought of as an extreme version of open set recognition where the \gls{dnn} is presented with a single class during training and is required to detect new incoming classes during testing or inference.

\noindent \textbf{Definition of Prior Shift:}~ Prior shift refers to a scenario where the marginal distribution of labels in source and target domain are different, i.e.,  $\mathcal{D}_S(Y) \neq \mathcal{D}_T(Y)$. On the other hand, the class conditional distribution of data given labels are assumed to be same, i.e., $\mathcal{D}_S(X|Y) = \mathcal{D}_T(X|Y)$. Prior shift might affect domain adaptation processes, also known as \gls{tta} in literature \cite{gong2022note,yuan2023robust}. Although the marginal label distribution $\mathcal{D}_S(Y)$ is uniform in most of the cases, in \gls{tta} the data is observed as a small batch of samples at a particular time. Hence, we might observe dominance of certain labels based on the current scenario. This is a very common example of prior shift that might happen which is also termed as '\textit{correlated label distribution}' in \gls{tta} literature.  

\noindent \textbf{Definition on Generalized Label Shift.}~ Generalized label shift is a more challenging data drift that was first introduced in \cite{tachet2020domain}. It occurs  when both covariate shift $\mathcal{D}_S(X) \neq \mathcal{D}_T(X)$ and prior shift $\mathcal{D}_S(Y) \neq \mathcal{D}_T(Y)$ occurs simultaneously. Moreover, it is assumed that there exists some feature representation $Z = g(X)$ for which the conditional distributions based on labels both on source and target domain are equal, i.e.,  $\mathcal{D}_S(Z|Y) = \mathcal{D}_T(Z|Y)$. This situation might arise when a \gls{dnn} continuously experiences different data corruptions while also being adapted with a skewed label distribution.

\begin{figure}
    \centering
    \includegraphics[width=0.75\textwidth]{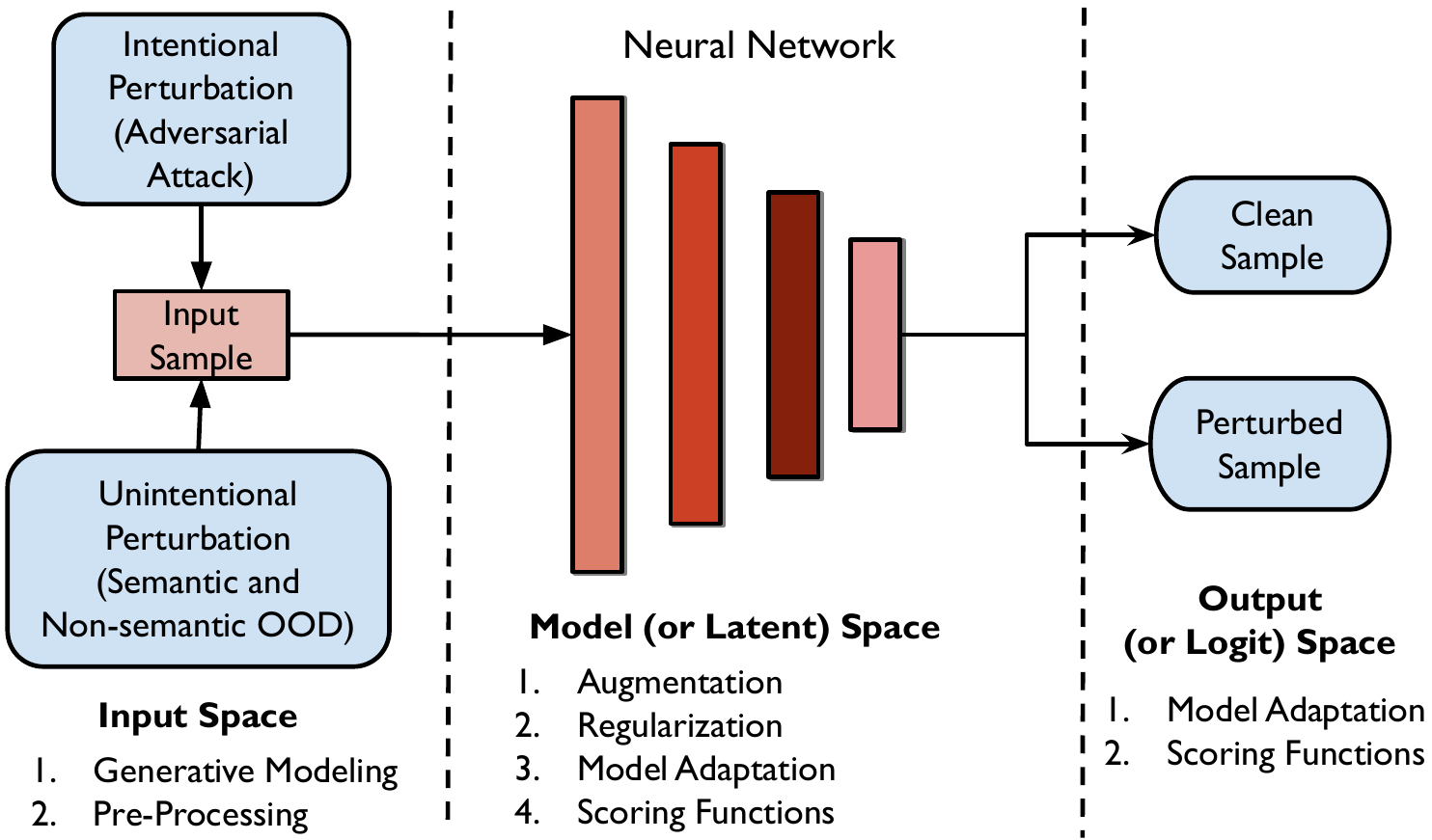}
    \vspace{0.2cm}
    \caption{Overview of existing approaches for \gls{dnn} resilience.}
    \label{fig:overview}
    \vspace{-0.3cm}
\end{figure}

\section{A Unified Vision for Intentional and Unintentional Perturbation}\label{sec:categorization}

There are two main approaches to ensure resilience of \glspl{dnn}. The first is by training the \gls{dnn} to be inherently resilient. Examples include adversarial training \cite{zhang2019you, jin2023randomized, jia2022adversarial} to avoid intentional perturbations, and training with different augmentation schemes \cite{hendrycks2021many,hendrycks2019augmix} for unintentional perturbations. The second way is to augment the \gls{dnn} with another binary classifier that differentiates between perturbed and unperturbed samples. Ideally, the auxiliary classifier would be able to differentiate between unperturbed and perturbed inputs for both intentional and unintentional perturbation. In reality, the detection of intentional and unintentional perturbation have developed as two separate domains. In the following, we highlight the connection among the methods for detection of intentional and unintentional perturbations. As such, we have categorized the approaches from the perspective of input space, latent (i.e., model) space, and  logit (i.e., output) space. Figure \ref{fig:overview} overviews our categorization, which is detailed below.\smallskip

\noindent \textbf{Input Space:}~Prior work has attempted to model the input distribution to detect intentional and unintentional perturbations. For example,  \cite{zhang2023detecting} proposed to model adversarially perturbed inputs with a diffusion model, which is also known as \textit{generative modeling}. The authors propose the \textit{expected perturbation score} which is the expected value of the gradient of the distribution of the output of the diffusion process at a given time, and show that the distribution of the expected perturbation score is different for the unperturbed and intentionally perturbed samples. For unintentional perturbation detection, the authors in \cite{serra2019input} propose a technique to learn the input distribution using generative models and corrects for the bias for complexity of the input samples.  \smallskip

\noindent\textbf{Model (or Latent) Space:}~Most of the existing work either adapts the model space or create scores based on the latent representations. The works in this space can befurther categorized into four types: (i) augmentation, (ii) regularization, (iii) model adaptation, and (iv) scoring functions. Some prior work has used \gls{ood} samples available in the wild to train the \gls{dnn} to distinguish between the \gls{id} and \gls{ood} samples by regularizing the loss function. The first work in this direction was \cite{hendrycks2019oe}, which was later improved in \cite{zhu2023diversified, wang2023outofdistribution} by improving the sampling of the \gls{ood} data as well as the regularization methodology. Prior work in \gls{ood} detection adapts the model to better differentiate between the \gls{id} and \gls{ood} inputs. For example, \cite{djurisic2022extremely, sun2021react, du2022siren} shape the model features during test-time for better detection of the \gls{ood} inputs. The work in \cite{lu2017safetynet} discretizes the output of a layer of the model into binary or quaternary codes and detects intentional perturbation by observing deviations from the codes produced by unperturbed samples. In addition, the authors in \cite{deng2021libre} focus on intentional perturbation detection and design a partial Bayesian network by learning the likelihood of the weights for the last few layers. This network is then used to measure the uncertainty associated with the input. A vast body of work in \gls{ood} crafts different scoring functions based on the latent representation of the \gls{dnn}. For example \cite{lee2018simple, ahn2023line, olber2023detection, deepneighborsun2022} craft scores using intermediate layer activations, while \cite{lee2018simple} also uses the score for detection of intentionally perturbed inputs.\smallskip

\noindent\textbf{Output (or Logit) Space:}~By output space, we indicate either the logit layer of a \gls{dnn} or its softmax probability distribution. The work in such output space has mostly focused on model adaptation and scoring functions. To the best of our knowledge, \cite{wei2022mitigating} is the only work we have found that focuses on the output space during training time. This work normalizes the output logits to mitigate the overconfidence issue of \glspl{dnn}. Conversely, \cite{roth2019odds} examines logits to detect the intentionally perturbed inputs and potentially correct the prediction. The work in \cite{hendrycks2016baseline} first used the maximum softmax probability to detect \gls{ood} samples, while \cite{liang2018enhancing, hsu2020generalized, Hendrycks2022ScalingOD, liu2023gen} studied the detection of \gls{ood} using score-functions derived from the output of a \gls{dnn}.

\section{Detection of Intentional Perturbation}
\label{sec:DAS}

Based on the classification in Section \ref{sec:categorization}, we describe existing work in perturbation detection as well as the related defense strategies. The detection of intentional perturbation  -- also known as adversarial perturbations in literature --  is mainly based on the hypothesis that adversarial samples reside in a manifold different than that of the unperturbed inputs \cite{Stutz2019CVPR, feinman2017detecting}. The key objective behind adversarial sample detection is to accurately detect and minimize the influence of adversarial samples on \glspl{dnn}. The general framework for this detection is: (a) characterizing the adversarial perturbation; (b) formulating a score function based on the characterization; and (c) deciding on a threshold to accept or reject a sample as adversarially perturbed. In this section, we first review some seminal work in this field and then provide key insights into recent work. We summarize all the reviewed work in Table \ref{tab:paper-list}.

\subsection{Use of Intermediate Representations}
\label{sec: perturbation}

One of the earliest approaches utilize the features from intermediate layers of a \gls{dnn} to detect adversarial samples. The authors of \cite{metzen2017detecting} train an adversary detector that receives inputs from the intermediate feature representations of a classifier. This detector aims to distinguish between samples from the original dataset and adversarial examples. The author considers two different scenarios: (i) a static scenario, where adversaries have access to the classification network and its gradients only; and (ii) a dynamic scenario, where adversaries have access to both the classification and the proposed detector network along with its gradients. The adversary detector network is trained in a supervised manner using training images and their corresponding adversarially perturbed images. The authors showed that a static detector cannot perform well against a dynamic adversary. 

To tackle dynamic adversaries, the author proposes a dynamic adversary training method inspired by the approach in \cite{goodfellow2014explaining}, where for each mini-batch the adversarial samples are computed on the fly. For each mini-batch, a dynamic adversary modifies a data point with a parameter $\sigma$ selected randomly from the range [0, 1], as it depends on the gradient of the detector which evolves over time. By training the detector, one implicitly makes it resilient to dynamic adversaries for various values of $\sigma$. Although it shows excellent results, it does not show robustness against random noise. One key takeaway from this work is that supervised training with specific attack strategy is unlikely to hold against adaptive attackers and a robust characterization of adversarial perturbation is required to be utilized by the detector so that it can generalize across different attack mechanisms instead of overfitting to a single type of attack. 

The detector sub-network proposed in \cite{metzen2017detecting} remains vulnerable to the adversarial samples that are not used during the detector training process. To address this issue,  \cite{lu2017safetynet} proposes a new approach called SafetyNet that relies on the hypothesis that adversarial attacks work by producing different patterns of activation in late-stage \glspl{relu} to those produced by natural examples. As a result, it focuses on discrete codes produced by the quantization of each \gls{relu} function at later stages of the classification network. SafetyNet consists of the original classifier along with an adversary detector that examines the internal state of the later layers in the original classifier.

\begin{landscape}
\label{tab:paper-list}
\begin{longtable}{|p{1.8cm}|p{.6cm}|p{2.8cm}|p{3.5cm}|p{7cm}|p{1cm}|p{0.6cm}|}
\caption{Studies on Adversarial Sample Detection Techniques. The metrics used refer to \textbf{A}: AUROC, \textbf{B}: Attack Failure Rate, \textbf{C}: Detection Accuracy/Detection Rate/Detection Success Rate, \textbf{D}: TPR(FPR@x), \textbf{E} : TPR, \textbf{F} : FPR, \textbf{G}: Reclassification Accuracy, \textbf{H}: Robust Accuracy, \textbf{I} : Robust Average Precision, \textbf{J} : Recovery Rate , \textbf{K} : Equal Error Rate, \textbf{L} : Label Change Rate, \textbf{M} : Detection Cost Function}

\\\hline

Work & Year & Task(s) & Dataset(s) & Attack(s) & Metrics & Code \\
\hline
\endfirsthead
\hline
Work & Year & Task(s) & Dataset(s) & Attack(s) & Metrics & Code \\
\hline
\endhead

Metzen et al. \cite{metzen2017detecting} & 2017 & Image Classification & CIFAR10\cite{krizhevsky2009learning}, 10-Class Imagenet\cite{deng2009imagenet} & \gls{fgsm}, \gls{bim}, DeepFool & C & \\
\hline 
Feinman et al. \cite{feinman2017detecting}   & 2017          & Image Classification                                                                               & MNIST\cite{deng2012mnist}, CIFAR10\cite{krizhevsky2009learning}, SVHN\cite{netzer2011reading}                                                                                                 & \gls{fgsm}, \gls{bim}, \gls{jsma}, \gls{cw}                                                                 & A                                                         & \href{https://github.com/rfeinman/detecting-adversarial-samples}{Code} \\
\hline
Lu et al. \cite{lu2017safetynet}        & 2017          & Image Classification                                                                               & CIFAR10\cite{krizhevsky2009learning}, Imagenet\cite{deng2009imagenet}                                                                                                    & \gls{fgsm}, DeepFool, \gls{ta}, \gls{bim},                             & C, E                                       & \href{https://github.com/Jianbo-Lab/ML-LOO.}{Code} \\
\hline
Grosse et al. \cite{grosse2017statistical}    & 2017          & Image Classification                                                                               & MNIST\cite{deng2012mnist}, DREBIN\cite{arp2014drebin}, MicroRNA\cite{shimomura2016rna}                                                                                              & \gls{fgsm}, \gls{jsma}, \gls{svma}, \gls{dta},                                                                & C & \\
\hline
Ma et al. \cite{ma2018characterizing}        & 2018          & Image Classification                                                                               & CIFAR10\cite{krizhevsky2009learning}, MNIST\cite{deng2012mnist}, SVHN\cite{netzer2011reading}                                                                                                 & \gls{fgsm}, \gls{bim}, \gls{jsma}, \gls{cw}                                                    & A, B                                    & \href{https://github.com/xingjunm/lid_adversarial_subspace_detection}{Code} \\
\hline
Zheng et al. \cite{Zheng2018}     & 2018          & Image Classification                                                                               & MNIST\cite{deng2012mnist}, F-MNIST\cite{xiao2017fashion}                                                                                                       & \gls{fgsm}, DeepFool                                                                      & A, E                                                    &  \\
\hline
Tao et al. \cite{tao2018attacks}       & 2018          & Facial recognition                                                                                 & VGGFace\cite{cao2018vggface2}, LFW\cite{huang2008labeled}, CelebA\cite{cao2018celeb} & \gls{fgsm}, \gls{bim}, \gls{cw}, Patching Attack, Glasses                           & F, C                                       & \href{https://github.com/AmIAttribute/AmI}{Code} \\
\hline
Lee et al. \cite{lee2018simple}       & 2018          & Image Classification                                                                               & CIFAR10\cite{krizhevsky2009learning}, CIFAR100\cite{krizhevsky2009learning}, SVHN\cite{netzer2011reading}                                                                                              & \gls{fgsm}, \gls{bim}, DeepFool, \gls{cw}                                                             & A                                     & \href{https://github.com/pokaxpoka/deep_Mahalanobis_detector}{Code} \\
\hline
Roth et al. \cite{roth2019odds}      & 2019          & Image Classification                                                                               & CIFAR10\cite{krizhevsky2009learning}, Imagenet\cite{deng2009imagenet}                                                                                                    & \gls{pgd}, \gls{cw},                                                                            & C, G                  & \href{https://github.com/yk/icml19_public}{Code} \\
\hline
Wang et al. \cite{wang2019adversarial}      & 2019          & Image Classification                                                                                & CIFAR10\cite{krizhevsky2009learning}, MNIST\cite{netzer2011reading}                                                                                                       & \gls{fgsm}, \gls{jsma}, DeepFool, \gls{cw}, Black-Box                                                 & A, C, L            & \href{https://github.com/dgl-prc/m_testing_adversatial_sample}{Code} \\
\hline
Cohen et al. \cite{cohen2020detecting}     & 2020          & Image Classification                                                                               & CIFAR10\cite{krizhevsky2009learning}, CIFAR100\cite{krizhevsky2009learning}, SVHN\cite{netzer2011reading}                                                                                              & \gls{fgsm}, \gls{jsma}, DeepFool, \gls{cw}, \gls{pgd}, \gls{ead},  Adaptive                                       & A, B                                    & \href{https://github.com/giladcohen/NNIF_adv_defense}{Code} \\
\hline
Yang et al. \cite{yang2020ml}      & 2020          & Image Classification                                                                               & MNIST\cite{netzer2011reading}, CIFAR10\cite{krizhevsky2009learning}, CIFAR100\cite{krizhevsky2009learning}                                                                                             & \gls{fgsm}, \gls{cw}, \gls{jsma} & A, D             &  \\
\hline
Li et al. \cite{li2020investigating}        & 2020          & Speaker identification                                                            & Voxceleb1 \cite{nagrani2017voxceleb}                                                                                                          & \gls{bim}, \gls{jsma}                                                                          & K, M, C & \\
\hline
Shumailov et al. \cite{shumailov2020towards} & 2020          & Image Classification                                                                                & MNIST\cite{netzer2011reading}, FashinMNIST\cite{xiao2017fashion}, CIFAR10\cite{krizhevsky2009learning}                                                                                          & \gls{fgsm}, \gls{bim}, \gls{pgd}, \gls{cw}, Fully Decision Based Boundary Attack                       & C                                            & \\
\hline
Deng et al. \cite{deng2021libre}      & 2021          & Image Classification, Face Recognition, Object Detection & Imagenet\cite{deng2009imagenet}, LFW\cite{huang2008labeled}, CPLFW\cite{zheng2018cross}, CALFW\cite{zheng2017cross}, CFP\cite{sengupta2016cfp}, VGGFace2\cite{cao2018vggface2}, AgeDB-30\cite{moschoglou2017agedb}, COCO\cite{lin2014microsoft}                                                           & \gls{fgsm}, \gls{bim}, \gls{pgd}, \gls{mim}, \gls{cw}, \gls{dim}, \gls{tim}                                                   & A                              & \href{https://github.com/thudzj/ScalableBDL.}{Code} \\
\hline
Li et al. \cite{li2021detecting}        & 2021          & Image Classification                                                                                & CIFAR10\cite{krizhevsky2009learning}, Imagenet\cite{deng2009imagenet}                                                                                                  & Patch-PGD, Adversarial Patch, Adaptive Attack                                                & C                                                & \\
\hline

Mekala et al. \cite{mekala2021metamorphic}    & 2021          & Facial Recognition                                                                                & VGGFace2\cite{cao2018vggface2}, LFW\cite{huang2008labeled},                                                                                  & \gls{fgsm}, \gls{pgd}, \gls{cw}                                                                       & C                                           &  \\
\hline
Xu et al. \cite{pai2023facade}      & 2018          & Image Classification                                                                                                & MNIST\cite{deng2012mnist}, CIFAR10\cite{krizhevsky2009learning}, Imagenet\cite{deng2009imagenet}                                                                                                                  & \gls{fgsm}, \gls{bim}, \gls{cw}, \gls{jsma}, DeepFool                                                                                 & H                                                           &               \\
\hline
Picot et al. \cite{picot2023a}     & 2023          & Image Classification                                                                               & CIFAR10\cite{krizhevsky2009learning}, CIFAR100\cite{krizhevsky2009learning}                                                                                                    & \gls{fgsm}, \gls{bim}, \gls{pgd}, \gls{cw}, DeepFool, \gls{hop}, \gls{sa}, \gls{sta}, Adaptive                                 & A, D                                                 & \href{https://github.com/MarinePICOT/HAMPER}{Code} \\
\hline
Chyou et al. \cite{chyou2023unsupervised}     & 2023          & Image Classification                                                                               & CIFAR10\cite{krizhevsky2009learning}                                                                                                              & \gls{fgsm}, \gls{bim}, \gls{pgd}, DeepFool, Auto Attack                                               & A, E, F                                               & \href{https://github.com/CycleBooster/Unsupervised-adversarial-detection-without-extra-model}{Code} \\
\hline
Zhang et al. \cite{zhang2023detecting}     & 2023          & Image Classification                                                                               & CIFAR10\cite{krizhevsky2009learning}, Imagenet\cite{deng2009imagenet}                                                                                                    & \gls{fgsm}, \gls{pgd}, \gls{bim}, \gls{cw}, Auto Attack, \gls{ta}                               & A                                                         & \href{https://github.com/ZSHsh98/EPS-AD.git}{Code} \\
\hline
Tarchoun et al. \cite{tarchoun2023jedi}  & 2023          & Image Classification, Object Detection  & Imagenet\cite{deng2009imagenet},Pascal VOC 07\cite{pascal-voc-2007}, INRIA\cite{thys2019fooling}, CASIA\cite{song2022casia}                                      & Adversarial Patch, \gls{lavan}, YOLO, Naturalistic Patch                & H, I, J      &  \\
\hline
Sun et al. \cite{Sun2023}       & 2023          & Image Classification                                                                               & CIFAR10\cite{krizhevsky2009learning}                                                                                                              & \gls{fgsm}, \gls{bim}, DeepFool, \gls{jsma}, \gls{cw}                                                       & C                                        & \\
\hline

\end{longtable}
\end{landscape}

\begin{figure}
    \centering
    \subfloat[Two simple "far away" 2D manifolds.]{
        \includegraphics[width=0.33\textwidth]{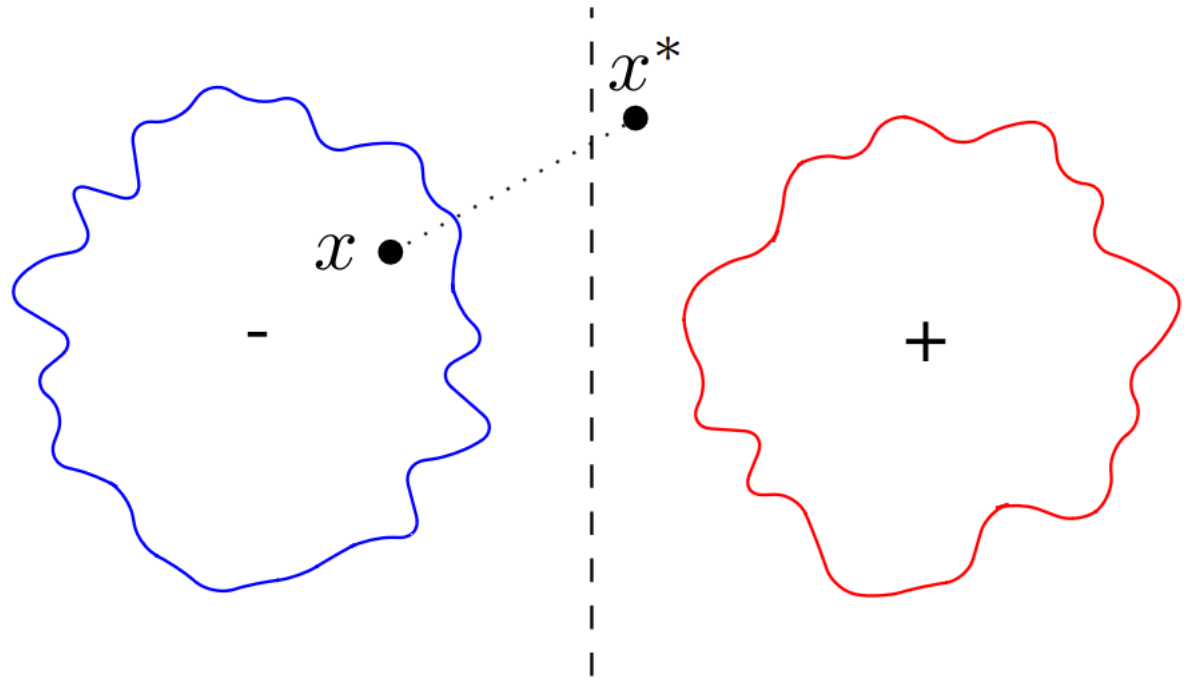}
        \label{fig:manifold_a}}
    \subfloat[One manifold having a "pocket".]{
        \includegraphics[width=0.33\textwidth]{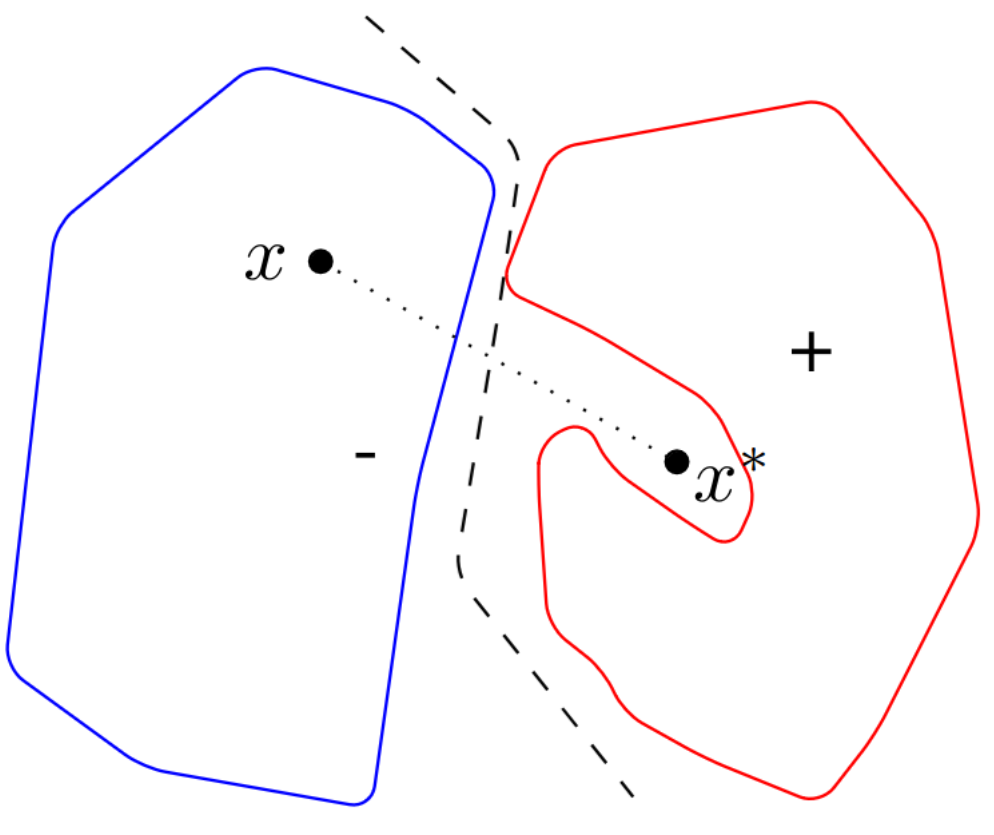}
        \label{fig:manifold_b}}
    \subfloat[Two simple "nearby" 2D manifolds.]{
        \includegraphics[width=0.33\textwidth]{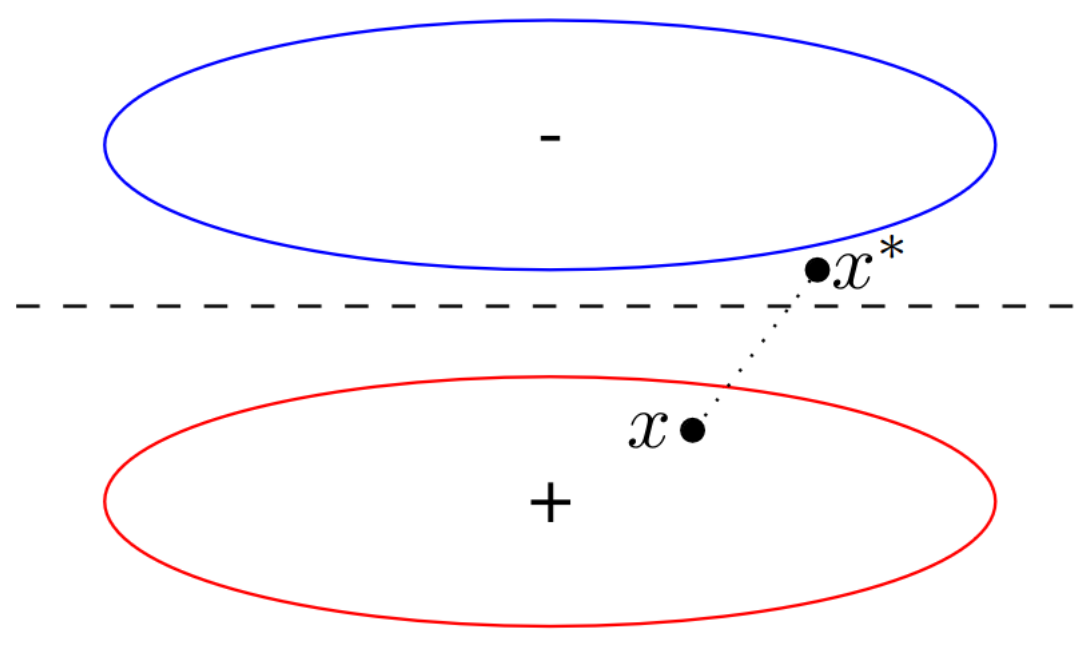}
        \label{fig:manifold_c}}
        \vspace{0.1cm}
    \caption{(a): The adversarial sample \(x^*\) is generated by shifting away from the 'negative' submanifold and crossing the decision boundary (black dashed line), but it remains distant from the 'positive' submanifold. (b): the 'positive' submanifold has a 'pocket' and the adversarial sample \(x^*\) lies in the pocket. (c): The adversarial sample \( x^*\) is close to both the decision boundary and both submanifolds \cite{feinman2017detecting}.} 
    \label{fig:manifold} \vspace{-0.5cm}
\end{figure}

The detector in SafetyNet employs a \gls{rbf-svm} to discern adversarial examples based on binary or quaternary codes representing activation patterns. For the code denoted by $b$, \gls{rbf-svm} uses the following formula to classify the samples:
\begin{equation}
    f(b) = \sum_{i}^{N}\alpha_{i}y_{i}\exp(-\parallel b-b_{i} \parallel^{2}/2\sigma^{2}) + c\;   
    \label{eq:safetyNet}
\end{equation}
The detector hardly generates any gradient unless the attacking code, denoted as $b$, closely resembles the code for a positive example $b_{i}$ when the variance of the kernel, $\sigma$ is very small. SafetyNet compels the attacker to solve a hard optimization problem that is discrete in nature. This work employs the idea of including non-differentiable operations so that the adversary cannot obtain the gradients necessary to compute adversarial perturbations. \vspace{-0.2cm}

\subsection{Manifold-based detectors}
\label{sec: manifold}

In \cite{feinman2017detecting}, the key intuition is that the adversarial samples lie on a different manifold than unperturbed samples. The authors argue that if after perturbation the data $x$ is transformed into $x^*$, it can leave the manifold $c_x$ in favor of sub-manifold $c_{x^*}$ in one of the following three ways illustrated in Figure \ref{fig:manifold_a}:
\begin{itemize}
    \item \(x^*\) is distant from the submanifold of \(c_{x^*}\) but closer to the classification boundary between \(c_x\) and \(c_{x^*}\);
    \item \(x^*\) lies closer to \(c_{x^*}\) submanifold but is still outside. On the other hand, \(x^*\) is distant from the classification boundary that separates the classes \(c_x\) and \(c_{x^*}\). As shown in Figure \ref{fig:manifold_b}, here one of the submanifolds has a pocket.
    \item \(x^*\) is close to the submanifold \(c_{x^*}\), but is still outside. In addition, \(x^*\) is close to the classification boundary which separates the classes \(c_x\) and \(c_{x^*}\). 
\end{itemize}

\noindent The authors estimate the manifold with \gls{kd} estimation. They do so with the output of the last hidden layer based on the hypothesis presented in \cite{gardner2015deep}, which states that the deeper layers of a \gls{dnn} offer more linear and 'unwrapped' manifolds compared to the input space. Given an input point $x$ and s set of training points $X_l$ having label $l$, the \gls{kde} $\hat{f}$ can be obtained as $\hat{f} = \frac{1}{\left|X_l  \right|}\sum_{x_i \in X_l}^{}k(x_i, x)$~
~, where $k(.,.)$ is the kernel function. The latter offers an indication of the distance between \( x\) and the submanifold for \(l\). For the point $x$, if the last hidden layer activation map is \(\phi(x)\), then the density estimate with predicted class $l$ is $\hat{k}(x, X_l) = \sum_{x_i \in X_l}^{}k_\sigma(\phi(x),\phi(x_i))$ 
where $\sigma$ is a tunable bandwidth. While this approach exhibits effective performance in the detection of adversarial samples that are located far away from the \(c_{x^*}\) submanifold, it performs poorly in scenarios where adversarial samples \( x^*\) lie in proximity to the \(c_{x^*}\) submanifold. As a result, in addition to the \gls{kde}, the author proposes \gls{bu} to identify low-confidence regions within the input space. The \gls{bu} shows considerably different distributions for normal and adversarial samples - supporting the intuition of different manifolds for adversarial examples. The drawback of their method is that the uncertainty estimate hinges on the use of "dropout" which limits its application. Also, the performance is poor on more challenging datasets and stronger attacks. \vspace{-0.3cm}

\subsection{\gls{lid}}
\label{sec: lid}

The objective of \gls{lid} \cite{ma2018characterizing} is to characterize the specific regions where adversarial examples may be located. Specifically, this work shows that the \gls{kd} approach adopted in \cite{feinman2017detecting}  -- which was based on the assumption that the adversarial subspaces are low probability regions -- fails to detect some forms of adversarial attack. As an alternative, the authors propose \gls{lid} to characterize the adversarial subspace. \gls{lid} represents the dimension of the data submanifold local to the data point $x$ under consideration. In connection to the classical expansion models, treating the probability mass as a proxy for volume may provide information about the dimensional structure of the data. \gls{lid} considers the \gls{cdf} of the number of data points encountered $F(d)$, where $d$ is a realization of the random variable $D$, i.e., the distance from data point $x$ to other data points. The \gls{lid} of $x$ at a distance $d$ can be defined as: 
\begin{equation}
    LID_{F}(d)\triangleq \lim_{\epsilon\rightarrow0}\frac{\ln(F((1+ \epsilon)\cdot d)/F(d))}{\ln(1+\epsilon)} = \frac{d \cdot F^{\prime}(d)}{F(d)}\;   
    \label{eq:lid}
\end{equation}
where $D$ is a positive random variable and the \gls{cdf} $F(d)$ of $D$ is continuously differentiable at distance $d>0$. The local dimension at $x$ in turn is defined as $LID_F = \lim_{d \rightarrow \inf} LID_F (d)$. The $LID_F$ quantifies how quickly the \gls{cdf} $F(d)$ grows with the distance $d$. It can be approximated by considering the distances between a point $x$ and its $k$ nearest neighbors within the dataset. The work in \cite{ma2018characterizing} hypothesizes that for the estimation of \gls{lid} of the adversarial samples, the nearest neighbors drawn should come not only from the manifold of the adversarial samples but also from the manifold of the normal samples, as the adversarial submanifold lies close to the data manifold. This will increase the dimension of the adversarial submanifold leading to higher value of \gls{lid}. This approach exhibits better generalization across different attacks than \gls{kd}. The authors train logistic regression model with \gls{lid} feature to discriminate between perturbed and unperturbed samples. The major drawback is that it fails against stronger attacks, which indicates that the characterization of the adversarial manifold with \gls{lid} is not universal for all adversarial perturbation. \vspace{-0.3cm}



\subsection{Interpretability-based Approaches}
\label{sec:interpretability}


Conversely from adopting adversarial sample training \cite{metzen2017detecting, feinman2017detecting}, the work in \cite{tao2018attacks} examines the adversarial samples from the \gls{dnn} interpretability point of view. Specifically, a novel adversarial sample detection named \gls{ami} has been proposed for face recognition. In this work, the main hypothesis is that adversarial samples utilize complex features extracted by the \gls{dnn} that are imperceptible to humans. As such, \gls{ami} initially extracts a set of neurons called \textit{attribute witness}, which are entangled with the face attributes. They substitute parts of the face from one image onto different images and look for unchanged neurons. This is named the \textit{Attribute Preservation} step. Another step is to substitute the same parts from different images onto a single image - generating versions of the image only part-substituted. The neurons which change in this case are likely to be attributed to that part, which is the \textit{Attribute Substitution} step. The common neurons obtained from attribute substitution and attribute preservation steps results in attribute witness neurons. The authors then construct an \textit{attribute-steered model} by increasing the values of the witness neuron and decreasing the values of non-witness neurons. For a given test input, the inconsistency observed between the two models indicates that the input is adversarial. The major drawbacks of this approach are the necessary manual feature inspection, absence of features in test image, and limited generalizability to setting other than face recognition. The work in \cite{fidel2020explainability} constructs \gls{shap}-based \cite{lundberg2017unified} feature attribution for the features of the penultimate layer. The knowledge base constructed from these \gls{shap} signatures for both natural and adversarial images are used to train a binary classifier. The key drawback of this approach is the requirement of the \gls{shap} signature from both natural and adversarially perturbed inputs. This creates a potential generalization problem for the binary classifier used for discriminating between the natural and adversarial inputs. \vspace{-0.5cm}

\subsection{Statistical Approaches}

The work in this category falls attempts to extract statistical information from different layers of the model to detect adversarial samples. The work in \cite{grosse2017statistical} proposes a statistical detection method based on \gls{mmd}. The authors hypothesize that only a limited number of samples is needed to  observe a measurable difference between normal and adversarial samples using a statistical test. They use a two-sample hypothesis test on the distribution of \gls{mmd} values to detect the difference between the normal samples and the adversarial samples. To detect single examples, they augment their classifier with an outlier class and train using the adversarial samples. The work in \cite{lee2018simple} proposed a \gls{gda}-based approach where the model features from different layers of the \gls{dnn} as class conditional multivariate Gaussian and calculate the confidence score for a sample as the  Mahalanobis Distance between the sample and the closest class conditional distribution. The authors extract such score from all the layers and integrate them using weighted averaging. Both \cite{grosse2017statistical, lee2018simple} utilize adversarial samples either to train their detector or find appropriate weights. These approaches are also model and attack specific and require separate detectors for different attack approaches. This severely limits their applicability.

The work in \cite{tarchoun2023jedi} approaches the detection of adversarial patches from a statistical perspective in the input space. The authors postulate that adversarial patches should contain a statistically higher amount of information, from an information theory perspective, compared to any random neighborhood from a natural image distribution. This leads to the proposal of \emph{Jedi}, which detects adversarial patches in images based on entropy thresholds. They use a 50-pixel $\times$ 50-pixel kernel and compute the entropy threshold dynamically based on the entropy distribution of the dataset and the image under consideration. The extracted high-entropy patch is passed through a sparse autoencoder for improved localization. Finally, they use coherence transport-based image inpainting \cite{bornemann2007fast} which aims at defusing the patch. They obtain high robust accuracy with respect to the baseline methods.\vspace{-0.3cm}

\begin{wrapfigure}{r}{0.35\textwidth}
    \centering
    \vspace{-0.5cm}
    \includegraphics[width=\linewidth]{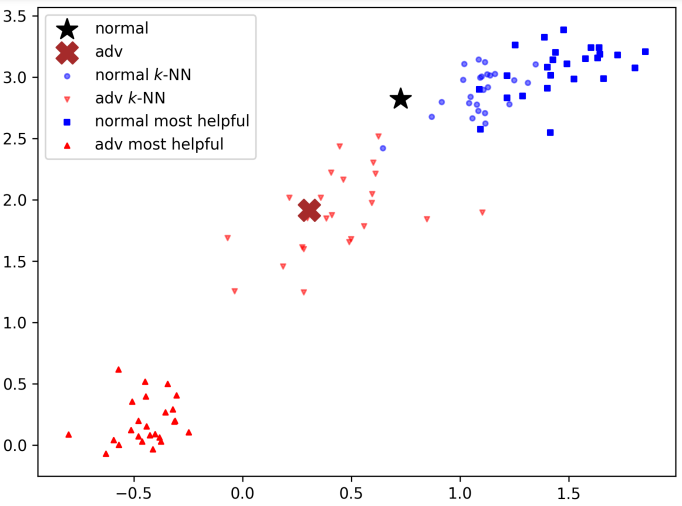}
    \caption{The embedding space of a \gls{dnn} \cite{cohen2020detecting}. \vspace{-0.5cm}}
    \label{fig:influence}
\end{wrapfigure}

\subsection{Influence-based approaches}
\label{sec:influence}

The work in \cite{cohen2020detecting} proposes a novel adversarial sample detection strategy by using an ``influence function'' \cite{koh2017understanding}. This approach can be employed by any pre-trained \gls{dnn}. The key intuition is that the training samples have close correspondence to the \gls{dnn} classification. When this relationship is disrupted, it strongly suggests the presence of an adversarial input. As such, the influence function measures the impact of training data in the decision-making of the \gls{dnn}. The influence of a training image $x$ on the loss of a specific test image $x_{test}$ can be measured as $I_{up, loss}(x, x_{test}) = -\nabla_{\theta}L(x_{test}, \theta)^TH_{\theta}^{-1}\nabla_{\theta}L(x, \theta)$ where $L$ is the loss function and $H$ is the Hessian of the machine learning model. The work also applies \gls{knn} classifier at the embedding space of the \gls{dnn} as the resemblance of the nearest neighbor in the embedding space also dictates the decision of the \gls{dnn}. The combination of the influence function and \gls{knn} classifier enables the detection of adversarial samples as the inference of a sample should be influenced most by the samples that are closest in the feature space. For adversarial samples, we observe the lack of this correlation. 

Figure \ref{fig:influence} illustrates this relation, where the black star and the brown X illustrate a normal and adversarial image from the CIFAR-10 validation set. We observe that the 25 nearest neighbors of the normal sample (blue circle) and the 25 helpful training samples (blue squares) for the normal sample lie very close in the \gls{pca} projected embedding space. A sharp contrast to this relation is observed in the case of the adversarial sample. Although this approach exhibits good performance in terms of generalization, it requires longer computation time as it calculates the influence function for the entire training dataset. The work in~\cite{zhang2023fast} masks the pixels of the input image and computes pixel level feature attribution by measuring the change in the output. The authors expedite the feature attribution to input space by using different sampling strategies. All these approaches assume access to the adversarial samples, which limit the utility of such approaches as they fail to detect stronger attacks. \vspace{-0.3cm}

\subsection{Other Notable Approaches}

\textbf{Input Space Approaches:}~\cite{zhang2023detecting} introduce a diffusion based perturbation method and derive \gls{eps}. They show that the distribution of \gls{eps} is different for normal and perturbed images. The difference in perturbation is measured using \gls{mmd}. This approach achieves an AUC of 1, and its performance does not degrade for unseen attacks. This is because \gls{eps} models the distribution of the input data itself instead of modeling the feature space. One drawback of the \gls{eps} score is that it cannot differentiate between adversarial perturbation and noise perturbation. The work in \cite{yang2020ml} proposes \gls{mlloo} feature attribution based detection of adversarial samples. The authors observe that feature attribution or mapping of importance of input features to the final prediction behaves differently for unperturbed and adversarially perturbed images. Adversarial perturbation disperses the feature attribution scores with significant deviation from the normal samples. Equipped with this observation, the authors use simple statistics to characterize the deviation of the adversarial samples as it progress through the \gls{dnn} and aggregate the statistics with a logistic regression model to differentiate between adversarial and normal samples. Although the authors utilize the leave-one-out feature attribution, their approach is generic to any feature attribution method. The work in \cite{zhang2018saliencydetection} trains a binary classifier using the saliency data by concatenating the saliency map to the raw image along the channel, while \cite{xu2018feature} proposes to squeeze input features using bit-depth reduction and spatial smoothing (both local and non-local variants). The work compares the output probability distributions on the original input and the feature-squeezed inputs using $L_1$ distance which changes significantly for adversarial inputs while normal samples show no change. This is because feature squeezing removes unimportant non-robust features improving robustness. \smallskip

\noindent \textbf{Model Space Approaches:}~The work in \cite{wang2019adversarial} proposed to modify the \gls{dnn} and has implemented a self-contained toolkit named \textit{mMutant} that integrates mutation testing and statistical hypothesis testing on \glspl{dnn}. The key observation is that the sensitivity of the mutation on \gls{dnn} is more acute for adversarial samples compared to unperturbed samples. If the \gls{dnn} is slightly altered, there is a greater chance that the mutated \gls{dnn} will alter the label of the adversarial sample than the unperturbed sample. The empirical investigation confirms this inherent sensitivity of the adversarial samples against a group of \gls{dnn} mutants in terms of \gls{lcr}. However, with the increase of mutation rate, the distance of \gls{lcr} between adversarial and normal samples decreases. This approach is also prone to the generation of some false positives during adversarial detection. Shumailov \textit{et al.} \cite{shumailov2020towards} proposed a mechanism called \gls{ctt}, which incorporated the \textit{Taboo Trap} detection, as well as numerical bound propagation. It prioritizes on finding the overexcited neurons being driven by adversarial perturbations outside of a predetermined range. The incorporation of numerical bound propagation on \gls{ctt} certifies the detection bounds on activation values of \gls{cnn} against specific input perturbation sizes. The authors proposed three variants of \gls{ctt} namely lite, loose, and strict. Although \gls{ctt}-lite does not require fine-tuning, its defense capacity is limited. Further optimization through fine-tuning results in \gls{ctt}-loose having most samples detected as adversarial. On the other hand, \gls{ctt}-strict guarantees detection of adversarial samples withing a specific range of $l_{\infty}$ bound. Deng \textit{et al.} \cite{deng2021libre} proposed \gls{libre} - a \gls{bnn} based approach that does not require training an extra model. The authors combined the \textit{expressiveness} of deep ensemble \cite{lakshminarayanan2017simple} and \textit{efficiency} of last layer Bayesian learning \cite{kristiadi2020being}. Conversely from using the entire \gls{dnn} for deep ensemble, they convert the last few layers -- for example, the last residual block of ResNet50 --  to a deep ensemble-like architecture. They also proposed to quantify the uncertainty with feature variance instead of softmax variance. As \gls{libre} does not use any adversarial samples for its training, it is adversarial attack agnostic and treats the adversarial samples as \gls{ood} data. This broadens its application scenario to more than classification with no modification which is not observed for the supervised detection methods.


Finally, in \cite{picot2023a} the authors proposed \gls{hamper} to characterize the adversarial samples. They calculated the class conditional half-space mass depth for a subset of \gls{dnn} layers and aggregate the scores as weighted sum. To tune the weight values, they utilized adversarial samples, and showed state-of-the-art performance in both attack-aware and blind-to-attack scenarios. It is shown that the last layers carry more important information for the detection of adversarial samples. \gls{hamper} has the advantage of being robust against adaptive attacks. However, this work does not provide any understanding as to why the half-space mass depth metric is able to capture the difference between the adversarial and normal samples. \smallskip

\noindent \textbf{Output Space Approaches:}~Prior work has attempted to characterize adversarial examples in the output space to facilitate their detection. I-defender \cite{Zheng2018} modeled the distribution of the output of the linear layers and shows that the distributions are different for the normal and adversarially perturbed images. The authors used a mixture of Gaussian models to approximate the \gls{ihsd} for each class. If the class conditional probability is lower than a threshold for a sample, it is detected as adversarially perturbed. However, this approach performs poorly under moderate to strong attacks. Roth \textit{et al.} \cite{roth2019odds} proposed a statistical metric for the detection of adversarial samples based on expected value of perturbed log-odds. They showed that the robustness properties of perturbed log-odds statistics are different for natural and adversarial samples. The idea is that geometrically optimal adversarial manipulations are embedded into a cone-like structure they call "\textit{Adversarial Cone}". They also reported an intriguing finding, i.e.,  that adversarial samples are much closer to the ground truth unperturbed class than any other class. Based on these observations, they proposed the maximum expected deviation of the perturbed log-odds from its expected value as an indicator of an adversarial sample. Chyou \textit{et al.} \cite{chyou2023unsupervised} proposed an unsupervised adversarial sample detection method without any extra model. They proposed new training losses to improve detection accuracy. The main idea is to remove unnecessary features for false outputs and strengthen the true outputs. This is achieved by forcing all the false raw outputs in a mini-batch to have a uniform distribution during training. By doing so, false outputs become adversarially robust and only true outputs can be attacked. Any attack on the true output changes the raw false output values triggering an adversarial detection. The proposed training loss keeps the accuracy of the original classification task almost the same, around 86\% when using ResNet18 architecture and CIFAR10. Although the authors used adversarial examples to determine their threshold, they show that the estimated threshold for their binary detector generalizes for other stronger attacks. The achieved accuracy on CIFAR10 dataset for Resnet architecture falls short of the reported accuracy of >90\% \cite{he2016deep} although this can be related to the training strategy. 


\section{Detection of OOD Samples}

Existing approaches for \gls{ood} detection mostly operate in the output space. Indeed,  it has been shown that working with activations of earlier layers does not provide much improvement in detector performance \cite{sun2021react,gradnorm2021}. In this section, we first provide the discussion of seminal work in \gls{ood} detection, and then focus on the state of the art approaches. Table \ref{tab:ood_det_summary} provides a summary of the surveyed work regarding \gls{ood} detection.  \vspace{-0.3cm}

\begin{table}
\small
\centering
\caption{Summary of the major \gls{ood} detection methods. Here, the mapping of alphabets to datasets is as follows A = TinyImageNet \cite{le2015tiny}, B = LSUN \cite{yu15lsun}, C = iSUN \cite{xu2015turkergaze}, D = Places365 \cite{zhou2017places}, E = CIFAR10 \cite{krizhevsky2009learning}, F = CIFAR100 \cite{krizhevsky2009learning}, G = Imagenet-1k \cite{deng2009imagenet}, H = SVHN \cite{netzer2011reading}, I = Textures \cite{cimpoi14describing}, J = Gaussian Noise, K = Uniform Noise, L = DomainNet \cite{peng2019moment}, M = iNaturalist \cite{vanhorn2018inaturalist}, N = ImageNet-21k \cite{ridnik2021imagenet21k}, O = MS-COCO \cite{lin2014microsoft}, P = OpenImages \cite{openimages}, Q = Pascal-VOC \cite{pascal-voc-2007}, R=BDD-100k \cite{yu2020bdd100k}, S = Flowers-102 \cite{Nilsback08}, T = Caltech256\cite{griffin2007caltech}}
\begin{tabular}{|p{1.3cm}|c|c|c|c|c|c|}
\hline
Work                                                   & Task(s)              & ID Dataset(s)                                              & OOD Dataset(s)                                                                                      & Base Model                                                                   & Metrics                                                                              & Code \\ \hline
Liang et al \cite{liang2018enhancing} & Image Classification & \begin{tabular}[c]{@{}c@{}}E, F\end{tabular} & \begin{tabular}[c]{@{}c@{}}A, B, C, J, K\end{tabular} & \begin{tabular}[c]{@{}c@{}}DenseNet-BC(k=12)\\ WideResNet-28-10\end{tabular} & \begin{tabular}[c]{@{}c@{}}FPR@95\%TPR\\ Detection Error\\ AUROC\\ AUPR\end{tabular} & \href{https://github.com/facebookresearch/odin}{code} \\ \hline

Liu et al \cite{liu2020energy} & Image Classification & \begin{tabular}[c]{@{}c@{}}E, F\end{tabular} & \begin{tabular}[c]{@{}c@{}}B(Crop), B(Resize), \\C, D, H, I\end{tabular} & \begin{tabular}[c]{@{}c@{}} WideResNet \end{tabular} & \begin{tabular}[c]{@{}c@{}}FPR@95\%TPR\\ AUROC\\ AUPR\\ \gls{id} Test Error\end{tabular} & \href{https://github.com/wetliu/energy_ood}{code} \\ \hline

Hsu et al. \cite{hsu2020generalized} & Image Classification & \begin{tabular}[c]{@{}c@{}}E,F,L \end{tabular} & \begin{tabular}[c]{@{}c@{}}A (crop), A (resize), \\B (Crop), B (Resize), \\C, H, L, J, K \end{tabular}& \begin{tabular}[c]{@{}c@{}}DenseNet-BC(k=12)\\ ResNet\\ WideResNet-28-10 \end{tabular} & \begin{tabular}[c]{@{}c@{}}TNR@95\%TPR\\ AUROC\end{tabular} & \href{https://github.com/wetliu/energy_ood}{code} \\ \hline

Huang et al. \cite{gradnorm2021} & Image Classification & \begin{tabular}[c]{@{}c@{}}E, F, G \end{tabular} & \begin{tabular}[c]{@{}c@{}}B (crop), C, D, H, I, M \end{tabular}& \begin{tabular}[c]{@{}c@{}}Google BiT-S\\ DenseNet-121 \\ResNet-20 \end{tabular} & \begin{tabular}[c]{@{}c@{}}FPR@95\%TPR\\ AUROC\end{tabular} & \href{https://github.com/deeplearning-wisc/gradnorm_ood}{code} \\ \hline

Sun et al. \cite{sun2021react} & Image Classification & \begin{tabular}[c]{@{}c@{}}E, F, G \end{tabular} & \begin{tabular}[c]{@{}c@{}} B (Crop), B(Resize), \\D, H, I, M\end{tabular}& \begin{tabular}[c]{@{}c@{}}ResNet-50\\ResNet-18 \end{tabular} & \begin{tabular}[c]{@{}c@{}}FPR@95\%TPR\\ AUROC \\AUPR\end{tabular} & \href{https://github.com/deeplearning-wisc/react.git}{code} \\ \hline

Wei et al. \cite{wei2022mitigating} & Image Classification & \begin{tabular}[c]{@{}c@{}}E, F \end{tabular} & \begin{tabular}[c]{@{}c@{}}B (crop), B (resize), \\C, D, H, I \end{tabular}& \begin{tabular}[c]{@{}c@{}}WideResNet-40-2\\ ResNet-34 \end{tabular} & \begin{tabular}[c]{@{}c@{}}FPR@95\%TPR\\ AUROC \\AUPR\end{tabular} & \href{https://github.com/hongxin001/logitnorm_ood}{code} \\ \hline

Djurisic et al. \cite{djurisic2022extremely} & Image Classification & \begin{tabular}[c]{@{}c@{}}E, F, G \end{tabular} & \begin{tabular}[c]{@{}c@{}}B (crop), B (resize), \\C, D, H, I, M \end{tabular}& \begin{tabular}[c]{@{}c@{}}DenseNet-101\\ ResNet-50\\ MobileNetV2 \end{tabular} & \begin{tabular}[c]{@{}c@{}}FPR@95\%TPR\\ AUROC \\AUPR\end{tabular} & \href{https://andrijazz.github.io/ash}{code} \\ \hline

Hendryks et al. \cite{Hendrycks2022ScalingOD} & \begin{tabular}[c]{@{}c@{}}Image Classification\\Multi-label Prediction\\Segmentation  \end{tabular} & \begin{tabular}[c]{@{}c@{}}N, O, Q\end{tabular} & \begin{tabular}[c]{@{}c@{}}Species(Subset of M)\\Subset of N  \end{tabular}& \begin{tabular}[c]{@{}c@{}}DenseNet-101\\ ResNet-50\\ MobileNetV2 \end{tabular} & \begin{tabular}[c]{@{}c@{}}FPR@95\%TPR\\ AUROC \\AUPR\end{tabular} & \href{https://github.com/hendrycks/anomaly-seg}{code} \\ \hline

Du et al. \cite{du2022vos} & Object Detection & \begin{tabular}[c]{@{}c@{}}Q, R \end{tabular} & \begin{tabular}[c]{@{}c@{}}O, P \end{tabular}& \begin{tabular}[c]{@{}c@{}}ResNet-50\\ RegNetX-4.0GF \end{tabular} & \begin{tabular}[c]{@{}c@{}}FPR@95\%TPR\\ AUROC \end{tabular} & \href{https://github.com/deeplearning-wisc/vos}{code} \\ \hline

Li et al. \cite{li2023rethinking} & Image Classification & \begin{tabular}[c]{@{}c@{}}E, F, G \end{tabular} & \begin{tabular}[c]{@{}c@{}}B, D, E, F, H, S, T \end{tabular}& \begin{tabular}[c]{@{}c@{}}ResNet-50\\ RegNetX-4.0GF \end{tabular} & \begin{tabular}[c]{@{}c@{}}AUROC \end{tabular} & \href{https://github.com/lijingyao20010602/MOOD}{code} \\ \hline

Hendrycks et al.$^*$ \cite{hendrycks2019oe} & Image Classification & \begin{tabular}[c]{@{}c@{}}A, D, E, F, H\end{tabular} & \begin{tabular}[c]{@{}c@{}}D, E, F, H \end{tabular}& \begin{tabular}[c]{@{}c@{}}PixelCNN++ \end{tabular} & \begin{tabular}[c]{@{}c@{}}FPR@95\%TPR\\ AUROC\\AUPR \end{tabular} & \href{https://github.com/hendrycks/outlier-exposure}{code} \\ \hline

Wang et al. \cite{wang2023outofdistribution} & Image Classification & \begin{tabular}[c]{@{}c@{}}E, F, G\end{tabular} & \begin{tabular}[c]{@{}c@{}}B (Crop), C, D, H, M \end{tabular}& \begin{tabular}[c]{@{}c@{}}WRN-40-2\\ResNet-50 \end{tabular} & \begin{tabular}[c]{@{}c@{}}FPR@95\%TPR\\ AUROC \end{tabular} & \href{https://github.com/QizhouWang/DOE}{code} \\ \hline

Zhu et al. \cite{zhu2023diversified} & Image Classification & \begin{tabular}[c]{@{}c@{}}E, F\end{tabular} & \begin{tabular}[c]{@{}c@{}}B (Crop), B(Resize), D, H, I \end{tabular}& \begin{tabular}[c]{@{}c@{}}WRN-40-2 \end{tabular} & \begin{tabular}[c]{@{}c@{}}FPR@95\%TPR\\ AUROC\\AUPR \end{tabular} & \href{https://github.com/tmlr-group/DivOE}{code} \\ \hline

Du et al. \cite{du2022siren} & Object Detection & \begin{tabular}[c]{@{}c@{}}Q, R\end{tabular} & \begin{tabular}[c]{@{}c@{}}O, P \end{tabular}& \begin{tabular}[c]{@{}c@{}}DDETR \end{tabular} & \begin{tabular}[c]{@{}c@{}}FPR@95\%TPR\\ AUROC \end{tabular} & \href{https://github.com/deeplearning-wisc/siren}{code} \\ \hline

Sun et al. \cite{deepneighborsun2022} & Image Classification & \begin{tabular}[c]{@{}c@{}}E, F\end{tabular} & \begin{tabular}[c]{@{}c@{}}B (Crop), C, D, H, I \end{tabular}& \begin{tabular}[c]{@{}c@{}}ResNet-18 \end{tabular} & \begin{tabular}[c]{@{}c@{}}FPR@95\%TPR\\ AUROC \end{tabular} & \href{https://github.com/deeplearning-wisc/knn-ood}{code} \\ \hline

Liu et al. \cite{liu2023gen} & Image Classification & \begin{tabular}[c]{@{}c@{}}E, F, G\end{tabular} & \begin{tabular}[c]{@{}c@{}}B (Crop), C, D, G, H, I, M, P \end{tabular}& \begin{tabular}[c]{@{}c@{}}BiT-S-R101$\times$1\\ViT-B/16\\RepVGG-B3\\ResNet-50-D\\DeiT-B/16\\Swin-B/4 \end{tabular} & \begin{tabular}[c]{@{}c@{}}FPR@95\%TPR\\ AUROC \end{tabular} & \href{https://github.com/XixiLiu95/GEN}{code} \\ \hline

\multicolumn{7}{l}{\small $^*$ The authors use alternately one dataset as \gls{id} and the rest as \gls{ood} dataset similar to leave-one-out approach}
\end{tabular}
\label{tab:ood_det_summary}
\end{table}

\subsection{Input Space Approaches} 

Among the few works in this space, Li \textit{et al.} \cite{li2023rethinking} forced the classifier to implicitly learn the \gls{id} data distribution instead of just learning features for classification. They applied a preprocessing step of masking and then learn to reconstruct the original image from the masked image. This calibrates the \gls{dnn} for \gls{id} and \gls{ood} samples. Gao \textit{et al.} \cite{gao2023diffguard} utilized the diffusion model to learn the distribution of \gls{id} data. They learned a \gls{ddim} conditioned to the semantic labels, while during inference time, they inverted the image to obtain the latent representation using the \gls{ddim} and them reconstruct it from the latents. \vspace{-0.3cm}

\subsection{Output Space Approaches}

ODIN \cite{liang2018enhancing} is among the earliest work in the domain of \gls{ood} detection in the output (or logit) space. It adopted the baseline \cite{hendrycks2016baseline} where the authors utilize the softmax probability score to distinguish between \gls{id} and \gls{ood} samples and improve it by incorporating temperature scaling and input pre-processing steps. ODIN \cite{liang2018enhancing} shows that by manipulating the temperature parameter $T \in \mathbb{R}_+$, it is possible to increase the separation between the \gls{id} and \gls{ood} examples. The resulting score function $S_{\hat{y}}(\mathbf{x};T)$ is given by Equation \ref{eq:odin}:
\begin{align}
\begin{split}
    S_i (\mathbf{x};T) = \frac{exp(f_i(x)/T)}{\sum_{j=1}^{j=N}exp(f_j(x)/T)} \\
    S_{\hat{y}}(\mathbf{x};T) = \mbox{max}_i\ S_i(\mathbf{x};T)
\end{split}
    \label{eq:odin}
\end{align}
Here, $f_i(\mathbf{x})$ denote the logit value corresponding to i-th class for \gls{dnn} $\mathbf{f}$. The authors complemented the temperature scaling by perturbing the input image before feeding it into the \gls{dnn}. The perturbation procedure was inspired by \cite{goodfellow2014explaining}, which adds the perturbation to decrease the softmax score but here the perturbation boosts the softmax score prediction. The perturbation follows $\Tilde{\mathbf{x}} = \mathbf{x} - \epsilon\mbox{sign}(-\nabla_{\mathbf{x}}\mbox{log}S_{\hat{y}}(\mathbf{x};T))$.


The authors explained the effect of temperature scaling using $U_1 = \frac{1}{1-N} \sum_{i \neq \hat{y}} [f_{\hat{y}}(\mathbf{x}) - f_i(\mathbf{x})]$ and $U_2 = \frac{1}{1-N} \sum_{i \neq \hat{y}} [f_{\hat{y}}(\mathbf{x}) - f_i(\mathbf{x})]^2$. They showed that \gls{id} data contains some classes that are similar to each other, resulting in a higher value of $U_2$ even when $U_1$ value for \gls{id} and \gls{ood} data is the same. By taking the Taylor approximation, the softmax function can be expressed as $S_{\hat{y}} = \frac{1}{N - (U_1 - \frac{U_2}{2T})/T}$. This suggests that for very high values of $T$, the softmax score is dominated by the term $U_1$ compensating the negative effect of $U_2$ on the detection performance. They argue that this makes the \gls{id} and \gls{ood} data more separable. The authors argue from the Taylor expansion of the log softmax of the perturbed input $\Tilde{x}$ that \gls{id} images have a larger norm of the gradient of the score function compared to the \gls{ood} images which results in a higher score function values after perturbation.
\begin{equation}
    \mbox{log}S_{\hat{y}}(\mathbf{\Tilde{x}};T) = \mbox{log}S_{\hat{y}}(\mathbf{x};T) + \epsilon ||\nabla_{\mathbf{x}}\mbox{log}S_{\hat{y}}(\mathbf{x};T)||_1 +o(\epsilon)
    \label{eq:odin-log-softmax}
\end{equation}
The key observations are  (i) \glspl{dnn} produce outputs with larger variance for \gls{id} examples; and (ii) \glspl{dnn} have larger value of the gradients of the log-softmax score when applied to \gls{id} images.


Generalized ODIN \cite{hsu2020generalized} improved \cite{liang2018enhancing} without utilizing any \gls{ood} samples. The authors point out the limitation of the softmax classifier, as it is an approximation of the indicator function it gives a categorical distribution rather than a continuous distribution over the classes. To circumvent this limitation, the authors reformulated the posterior probability by incorporating domain of the input as a random variable, changing it from $p(y | \mathbf{x})$ to $p(y | \mathbf{x}, d_{in})$, where $y, \mathbf{x}, \mbox{and}\ d_{in}$ represent the output, the input and the input domain. This lead to the decomposition of the posterior as $p(y | \mathbf{x}, d_{in}) = \frac{p(y, d_{in} | \mathbf{x})}{p(d_{in}, \mathbf{x})}$. Due to the lack of out-of-domain knowledge, the authors proposed to utilize the prior knowledge of the dividend/divisor structure of the posterior to provide the classifier capacity to decompose the confidence of the class probability. They modeled the logits $f_i(\mathbf{x})$ as $f_i({\mathbf{x})} = \frac{h_i({\mathbf{x}})}{g(\mathbf{x})}$. While the work shows improvement over \cite{hendrycks2016baseline,liang2018enhancing,lee2018simple}, its key limitation is that it needs to change the structure of the \gls{dnn} and perform training to learn $h_i(\mathbf{x})$ and $g(\mathbf{x})$.

\begin{wrapfigure}[12]{r}{0.65\textwidth}
    \vspace{-0.5cm}
    \centering
    \includegraphics[width=0.65\textwidth]{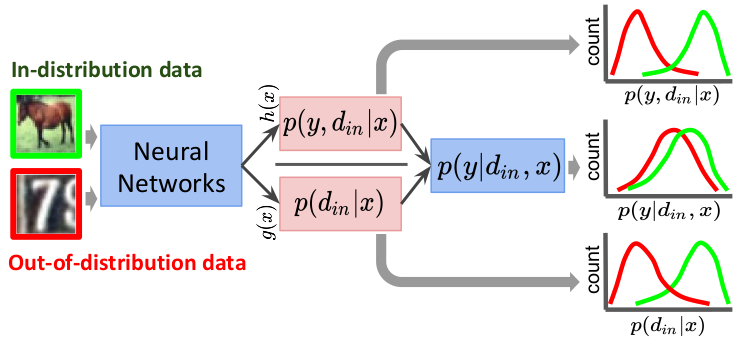}
    \caption{Decomposing the logit values into dividend/divisor structure. g(x) allows incorporating domain knowledge into training and disentangles the probability over the domain and probability over the classes. \cite{hsu2020generalized}}
    \label{fig:enter-label}
\end{wrapfigure}

Liu \textit{et al.}~\cite{liu2023gen} proposed a generalized entropy-based approach. Specifically, the authors used the generalized entropy family defined by $G(\mathbf{p}) = \sum_i p_i^\gamma(1-p_i)^\gamma$, where $\mathbf{p}$ denotes a categorical probability distribution. This score function, coupled with truncation of very small probabilities, showed competitive result for state-of-the-art models like Swin Transformer and BiT-S R101$\times$1 on ImageNet benchmark. The work in \cite{Hendrycks2022ScalingOD} addressed the fact that OOD detection schemes underperform in real-world settings. This work introduces a new dataset called \textit{Species} consisting of 700,000 images and over 1,000 anomalous species to test \gls{ood} detection performance on the ImageNet dataset. They proposed to use the negative of the maximum of the logit values (Max Logit) of a \gls{dnn} as the score to distinguish between \gls{id} and \gls{ood} samples and established a baseline for large-scale setting. They also introduced new benchmarks for anomaly segmentation and \gls{ood} detection in multi-label prediction. Decoupling max-logit \cite{zhang2023decoupling} decouples the magnitude and the direction of the logit vector. This decomposes the max-logit score into product of max cosine (cosine of the angle between the feature vector and respective weights for each class) and max norm (norm of the product between feature vector and respective weights for each class). While max-cosine outperforms max-logit score consistently, max norm falls short by a large margin. 

GradNorm \cite{gradnorm2021} is among the approaches using the gradients of the parameters of the \gls{dnn}.  The approach utilizes a label-agnostic score function to formulate the \gls{ood} detection problem as a binary classification problem. For a \gls{dnn} parameterized with $\mathbf{w}$, GradNorm formulates the score function for an input $\mathbf{x}$ as $S(\mathbf{x}) = ||\frac{\partial D_{KL}(\mathbf{u} || softmax(f(\mathbf{x})))}{\partial\mathbf{w}}||_p$, where $\mathbf{u}$ denotes the uniform distribution and $D_{KL}$ denotes the KL-divergence. The intuition behind this formulation is that the prediction tends to concentrate around the target class for \gls{id} inputs which should result in a larger value of the KL divergence and its derivative. The parameter vector $\mathbf{w}$ contains the concatenated parameters from different layers in a single vector regardless of their original shape. The authors further show that the gradients from the last layer's parameters are sufficiently informative for distinguishing between the \gls{id} and \gls{ood} inputs. Through empirical results, they showed that the $L_1$ norm is the most effective for GradNorm. The authors attribute the improvement brought by GradNorm to the joint information from the feature space and the output space that is utilized in this framework. Specifically, they showed that the score function can be written as equation \ref{eq:gradnorm_decomposition}, where the $x_i$s are the input features to the logit layer, $C$ is the number of classes, $T$ is the temperature, and $f_j$ denotes the logit for class $j$. $U=\sum_{j=1}^{m}|x_i|$ and $V$ represent the information about the feature space and output space respectively.

\vspace{-0.5cm}
\subsection{Approaches based on Energy Functions}

The work \cite{liu2020energy} provides a solution to \gls{ood} detection problem in the logit space. It connects the posterior probability in Gibb's distribution and the posterior probability of the softmax layer and proposes the equivalent of Helmholtz energy function for the softmax function as the score function for discriminating between \gls{id} and \gls{ood} samples. The key observation is the similarity between Gibb's density function in Equation \ref{eq:energy_gibbs} and the softmax function  in Equation \ref{eq:energy_softmax}. By connecting these two equations, the energy for a specific sample is defined by $E(\mathbf{x}, y) = -f_y(\mathbf{x})$:
\begin{equation}
    p(y | \mathbf{x}) = \frac{e^{-E(\mathbf{x},y)/T}}{\int_{y'} e^{-E(\mathbf{x}, y')/T}} = \frac{e^{-E(\mathbf{x},y)/T}}{ e^{-E(\mathbf{x})/T}}
    \label{eq:energy_gibbs}
\end{equation}
\begin{equation}
    p(y | \mathbf{x}) = \frac{e^{f_y(\mathbf{x})/T}}{\sum_{i=1}^{i=K} e^{f_i(\mathbf{x})/T}}
    \label{eq:energy_softmax}
\end{equation}

This also leads to the equivalent of free-energy $E(\mathbf{x})$ for \gls{dnn} as $E(\mathbf{x};f) = -T \sum_{i=1}^{i=K}e^{f_i(\mathbf{x})/T}$. The authors show that the Energy function is affinely related to the logarithm of the probability of the input $p(\mathbf{x})$. On the other hand, the logarithm of the softmax score or the softmax score itself is not related to $log\ p(\mathbf{x})$ linearly and depends on the maximum logit value. The authors also proposed an energy-based regularizer for training the \gls{dnn}, shown in Equation \ref{eq:energy_regularizer} and Equation \ref{eq:energy_regularizer_},  and demonstrated that it improves the performance of the detector. 
\begin{equation}
    min_{\theta}\ \mathbb{E}_{(\mathbf{x}, y)\sim \mathcal{D}_{in}^{train}} [-log F_y(\mathbf{x})] + \lambda L_{Energy}
    \label{eq:energy_regularizer}
\end{equation}
\begin{equation}
    L_{Energy} = \mathbb{E}_{(\mathbf{x}, y)\sim \mathcal{D}_{in}^{train}} (max(0, E(\mathbf{x}_{in}) - m_{in}))^2 + \mathbb{E}_{(\mathbf{x}, y)\sim \mathcal{D}_{out}^{train}} (max(0, m_{in} - E(\mathbf{x}_{out})))^2
    \label{eq:energy_regularizer_}
\end{equation}
Despite having the advantage of being easy in nature, the energy-based \gls{ood} detection method requires access to the \gls{ood} distribution either to determine the threshold or to train the network with regularization. \vspace{-0.5cm}

\subsection{Approaches based on Activation Shaping}

\begin{wrapfigure}{l}{0.35\textwidth}
    \centering
    \includegraphics[width=\linewidth]{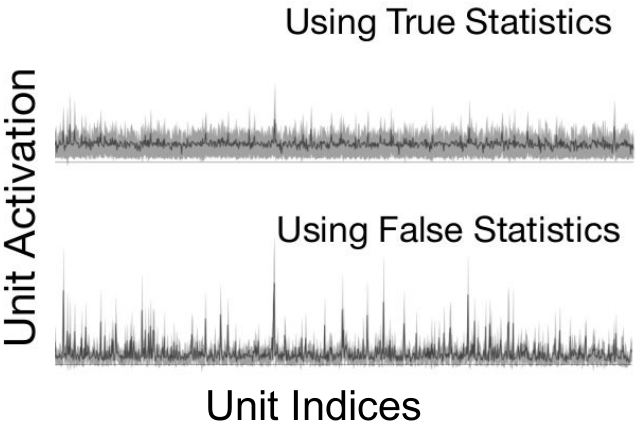}
    \caption{The per-unit activations for the penultimate layer for OOD data (iNaturalist) are considerably different when using \textit{true} (top) statistics as opposed to \textit{mismatched} (bottom) BatchNorm statistics \cite{sun2021react}.\vspace{-0.5cm}}
    \label{fig:react_bn_layer}
\end{wrapfigure}

These approaches assume that the features of a \gls{dnn} affect the output differently for \gls{id} and \gls{ood} samples. As such, they rely on some predefined transformation of the activation values of intermediate layers of \gls{dnn} to differentiate between \gls{id} and \gls{ood} samples. ReAct \cite{sun2021react} is a rectification operation on the activation in model space to facilitate the detection of \gls{ood} samples by making \gls{id} and \gls{ood} samples more separable. The activations from the penultimate layer $h(\mathbf{x})$ of the \gls{dnn} are truncated using ReAct operation given by $\Bar{h}(\mathbf{x}) = \mbox{ReAct}(h(\mathbf{x}; c))$, where $\mbox{ReAct}(x;c)=min(x,c)$.

The rectified activations are used to obtain the model output as $f^{ReAct}(\mathbf{x};\theta)=\mathbf{W}^T\Bar{h}(\mathbf{x})+\mathbf{b}$. These outputs can be used with any score function for \gls{ood} detection. The detection performance depends on the threshold value $c$. Through empirical study, the authors showed that setting the value of $c$ to the 90th percentile of the activations works best for ReAct. The authors theoretically showed that ReAct suppresses the activations more for the \gls{ood} samples than \gls{id} samples. This translates to a larger reduction in the output value for the \gls{ood} samples making them  separable from the \gls{id} samples. When batch-normalization statistics calculated for \gls{id} data are applied to \gls{ood} data, significantly different activation patterns emerge as shown by the authors in Figure \ref{fig:react_bn_layer}. The approach is also shown to work for \gls{dnn} architectures using normalization techniques other than batch-normalization (e.g. weight normalization, group normalization).


\noindent Activation Shaping (ASH) \cite{djurisic2022extremely} modifies the activations of a \gls{dnn} in a post-hoc manner to facilitate the detection of \gls{ood} samples using existing scoring methods. The work assumes that modern overparameterized \gls{dnn} produce redundant representations for the task it is trained for. Therefore, the representation can be greatly simplified while preserving performance and providing additional benefits in \gls{ood} detection. Based on this hypothesis, the authors propose three variants of the ASH algorithm where they set values smaller than p-th percentile of the representation to 0 as a form of activation pruning. The first variant, \textbf{ASH-P}, keeps the un-pruned activations unchanged. In the second variant, \textbf{ASH-B}, the un-pruned activations are assigned a value such that their total value equals that of the total value of the original activations. The third variant, \textbf{ASH-S}, calculates $s_1$ and $s_2$, the sums of the activations before and after pruning, and scales the unpruned values with $exp(s_1/s_2)$. They apply this activation shaping to the penultimate layer of the \gls{dnn}. The effect of the choice of the value of $p$ for setting the pruning threshold is dependent on the architecture and dataset and takes values in the range of [65, 95]. While this method is relatively simple, it integrates the detector into the architecture impacting the performance on \gls{id} data. The unique perspective of this method indicates that sparsity can be beneficial to \gls{ood} detection.
\begin{equation}
    S(\mathbf{x}) = \frac{1}{CT}(\sum_{i=1}^{m}|x_i|)(\sum_{j=1}^{C}|1 - C.\frac{e^{f_j/T}}{\sum_{j=1}^{C}e^{f_j/T}}|) \overset{\Delta}{=} \frac{1}{CT}U \cdot V
    \label{eq:gradnorm_decomposition}
\end{equation}
This approach is label-agnostic, \gls{ood} data agnostic, and can be utilized using back-propagation. Conversely, other approaches that are comparable in accuracy require considerable computational overhead and also require access to \gls{ood} data \cite{lee2018simple}. The limitation of this work lies in the assumption that the softmax output probability would be more uniform for \gls{ood} data. It has been shown by \cite{nguyen2015deep} that \gls{dnn} can assign a high probability to a specific class even when presented with random noise. This limitation is reflected in the results presented in the work where it is shown that GradNorm still reports \gls{fpr} of 43.16\% at 95\% \gls{tpr}.

Similar to \cite{djurisic2022extremely}, the work in \cite{ahn2023line}  suppresses the irrelevant activations. However it also suppresses weights based on Shapley value \cite{shapley1953value}. The impact on the performance for \gls{id} requires further study in such cases where the \gls{dnn} architecture is changed.  The work \cite{zhu2022boosting} takes a similar approachwhere they assume that the deep features follow a Gaussian distribution. They define features that fall into the high-probability region as \textit{typical features} and they estimate the typical features with the batch-normalization layer statistics. For a batch-normalization layer with mean $\mu$ and standard deviation $\sigma$, the features that are within $\lambda\sigma$ from $\mu$ are estimated to be the typical features. The features are then clamped to have values between $\mu-\lambda\sigma$ and $\mu+\lambda\sigma$ meaning that only the typical features can pass unperturbed. 

\gls{vra} \cite{xu2023vra} adopts a piece-wise activation shaping function similar to \cite{zhu2022boosting}. The difference is that they clamp the lower values to zero and shift the intermediate values by a fixed amount $\gamma$. The inspiration to shift the intermediate values comes from the variational formulation of the activation-shaping function, which shows the need to suppress abnormally low and high values and boosting intermediate feature values. Both \cite{zhu2022boosting} and \cite{xu2023vra} are  inspired by ReAct \cite{sun2021react}. The recent work \cite{zhao2024towards} formulates the design of feature-shaping functions as an optimization problem and shows that the previous approaches \cite{sun2021react,djurisic2022extremely} approximate the optimal solution. The authors provide two versions of the optimization problem. The first one utilizes the \gls{ood} data and the second one is \gls{ood} data-free approach. For the \gls{ood} data-free approach, the authors also provide a closed-form solution.

\vspace{-0.4cm}
\subsection{Approaches based on Mitigating Overconfidence in Prediction}

LogitNorm \cite{wei2022mitigating} observes that the norm of the logit vector increases as the training progresses. As up-scaling the logits increases the maximum softmax probability, to normalize the logit values to address the issue. Their key idea is to decouple the influence of the norm of the output from the training objective and optimization step. They decomposed the logit vector $\mathbf{f}$ into its norm and directions as given by $\mathbf{f} = ||\mathbf{f}|| \cdot \hat{\mathbf{f}}$. They proposed to constrain the logit norm to a fixed value $\alpha$, leading to the constrained optimization problem given by Equation \ref{eq:logit_norm_optimization}:
\begin{equation}
\begin{split}
    \mbox{minimize}\ \mathbb{E}_{\mathcal{P}_{\mathcal{XY}}}[\mathcal{L}_{CE}(f(\mathbf{x;\theta}, y))]\\
    \mbox{subject to}\ ||f(\mathbf{x;\theta})||_2 = \alpha
\end{split}
    \label{eq:logit_norm_optimization}
\end{equation}
They showed that this constrained optimization does not work in practice and propose to use the logit vector normalized with its $L_2$ norm instead. This leads to the optimization objective given by Equation \ref{eq:logit_norm_objective}:
\begin{equation}
    \mathcal{L}_{logit-norm}(f(\mathbf{x}; \theta), y) = - \mbox{log}\frac{e^{f_y/(\tau ||\mathbf{f}||)}}{\sum_{i=1}^{k}e^{f_i/(\tau ||\mathbf{f}||)}}
    \label{eq:logit_norm_objective}
\end{equation}
where the temperature parameter $\tau$ modulates the logit-norm. In this case, the compounded temperature parameter $\tau ||\mathbf{f}||$ shows dependence on the input as $\mathbf{f} = f(\mathbf{x};\theta)$. The authors argued that the input-dependent temperature scaling provides superior performance in terms of \gls{ood} detection and calibration of the output. While providing a simple fix to the overconfidence issue of the \gls{dnn}, LogitNorm approach has the drawback that the output distribution gets shifted due to the normalization. The paper does not provide any theoretical guarantees for improvement in calibration performance. While this approach can improve the performance of different scoring functions, it requires modification of the training objective. Hence,  it cannot be applied to pretrained \glspl{dnn}.

\gls{mood} \cite{li2023rethinking} shows that the overconfident prediction in the \gls{ood} samples is due to incomplete learning of the \gls{id} features. As such, it can be mitigated by using masked image modeling (MIM) as a pretext task. The authors argued that using the reconstruction of the masked image as the pretext task forces the \gls{dnn} to learn the \gls{id} data distribution instead of just patterns for classification and this improves the \gls{ood} classification performance. The Mahalanobis distance \cite{lee2018simple} proves to be the most effective. This suggests a Gaussian distribution of features and well-separated class boundaries. The overall approach of \gls{mood} consists of four steps - a) pretraining the Masked Image Modelling \gls{vit} on ImageNet-21k., b) intermediate fine-tuning of \gls{vit} on ImageNet-21k, c) fine-tuning of pre-trained \gls{vit} on the \gls{id} dataset, and d) extracting features from the pre-trained \gls{vit} and using Mahalanobis distance metric to detect \gls{ood} samples. Despite the superior performance shown by \gls{mood}, a shortcoming is that only the \gls{auroc} is used as the evaluation metric.

The work in \cite{du2022vos} addresses the problem of lack of supervision from \gls{ood} data during training resulting in overconfident predictions on data from unknown domains. The authors propose an unknown-aware training approach for object detection tasks without accessing \gls{ood} data. Conversely, they create virtual outliers to regularize the model. They model the feature representation $h(\mathbf{x,b})$ of object instances $\mathbf{x}$ ($\mathbf{b}$ is the bounding box) with class-conditional multivariate Gaussian distribution having mean $\mu_k$, and covariance matrix $\Sigma$. After fitting the features from the penultimate layer of the network to the class conditional Gaussian distributions, they sample feature instances ($\mathbf{v}$) from these distributions to use for regularizing the \gls{dnn}. For regularization, they use an uncertainty loss $\mathcal{L}_{uncertainty}$ as follows:
\begin{equation}
    \mathcal{L}_{uncertainty} = \mathbb{E}_{\mathbf{v} \sim \mathcal{V}}[-\mbox{log} \frac{1}{1 + exp^{-\phi(E(\mathbf{v};\theta))}}] + \mathbb{E}_{\mathbf{x} \sim \mathcal{D}}[-\mbox{log} \frac{exp^{-\phi(E(\mathbf{x};\theta))}}{1 + exp^{-\phi(E(\mathbf{x};\theta))}}],
    \label{eq:vos_loss}
\end{equation}
where $\mathcal{V}$ and $\mathcal{D}$ denote the virtual outlier and training datasets respectively, while $\phi(\cdot)$ is a non-linear MLP function that allows learning flexible energy surface. This work provides a flexible approach for use in \gls{ood} detection for object detection tasks, which is an underexplored area. However, the performance still needs to improve significantly. Indeed, the reported \gls{fpr} gets close to 50\%, which may be unacceptable in real-world tasks. \vspace{-0.3cm}

\subsection{Approaches based on Training with Regularization}

Some prior work assumes access to \gls{ood} data -- also referred to as  surrogate \gls{ood} data -- that can be used at training time. Specifically, these approaches propose a regularization term to be added to the training loss that helps improve detection of \gls{ood} samples. The key issue with surrogate OOD data is that the OOD distribution may not be fully characterized. As a result, these methods perform poorly on unseen \gls{ood} distributions. \gls{oe} \cite{hendrycks2019oe} is the first to propose to use such surrogate data. They assumed to have training OOD data that is disjoint from the test \gls{ood} data and proposed a generalized training scheme using the regularizer $\mathcal{L}_{OE}$, which is task-dependent. The authors set $\mathcal{L}_{OE}$ to cross-entropy loss between the \gls{dnn} output and uniform distribution for supervised learning tasks. If the task is of density estimation and without any label, then $\mathcal{L}_{OE}$ is set to margin ranking loss between output of \gls{dnn} $f$ - $f(x)$ and $f(x')$, where $x$ and $x'$ are sampled from \gls{id} and surrogate \gls{ood} distribution respectively. 

\gls{doe} \cite{wang2023outofdistribution} proposes an improvement over \gls{oe} by making it distribution-agnostic. \gls{doe} is composed of \gls{ood} generation using model perturbation and worst OOD regret (WOR) based training. The perturbation is done with a matrix multiplying the \gls{dnn} parameters, which leads to features resembling ones sampled from the \gls{ood} distribution. The WOR score measures the worst performance of the \gls{ood} detector and helps discover the hardest \gls{ood} samples. The hardest \gls{ood} samples are simulated by perturbing the model parameters. \gls{divoe} \cite{zhu2023diversified} took a similar approach as \gls{doe}. However, instead of generating new \gls{ood} samples through model perturbation, they perturbed a portion of the surrogate \gls{ood} data and use a mix of the perturbed (that maximize $\mathcal{L}_{OE}$ loss) and unperturbed surrogate data to train the \gls{dnn}.   

The work by Choi \emph{et al.}~\cite{choi2023balanced} proposed a balanced energy regularization loss $L_{energy, bal}$, which is built upon the energy regularization loss $L_{energy}$ proposed in \cite{liu2020energy}. The authors fine-tuned the linear layers of a \gls{dnn} with auxiliary samples (i.e. \gls{ood} samples) and used the proposed loss as a regularization factor, which increases the energy margin between \gls{id} and \gls{ood} samples. The proposed loss accounts for the imbalance in the \gls{ood} data used for training the \gls{dnn}. However, such training requires access to \gls{ood} data and is computationally expensive. Another line of work considers adding regularizers to the loss function without using any surrogate \gls{ood} data. Conversely, these approaches impose constraints on latent representations of \gls{id} data that \gls{ood} data are unlikely to achieve. For example, \cite{zaeemzadeh2021out} proposed to constrain the embedding generated by \gls{dnn} to a union of 1-dimensional subspaces where inputs from each class occupy one dimension. The authors achieve such embeddings by imposing a cosine similarity constraint between the embeddings and the weight vector corresponding to the class it belongs to. They also imposed orthogonality constraints among the weight vectors of different classes so as not to impact the \gls{id} accuracy. The \gls{ood} samples are  detected by estimating the probability of a test sample belonging to one of the classes using spectral discrepancy (angular deviation of the embedding from first singular vector corresponding to a class) measurement. If the probability of the sample belonging to any class is very low (ideally zero), then it is declared to be an \gls{ood} sample. 

Du \textit{et al.}~\cite{du2022siren} proposed \textit{Siren}, which added a projection head after the penultimate layer of the \gls{dnn}. Such projected features are modeled with a \gls{vmf} distribution \cite{mardia2009directional} and trained to align with class conditional priors. The score function is designed with maximum class conditional likelihood, while the regularizer used to train the projections ensures that the \gls{id} representations align with their class conditional priors, which is unlikely for \gls{ood} samples.  Finally, the work in \cite{ming2023exploit} proposed CIDER, which utilized the distance from class prototypes in a hyper-spherical embedding space. The loss function in CIDER is divided into two sections. The first one makes the embedding compact around each class prototype, while the second increases the angular distance among the class prototypes. The authors show that using the two part loss provides better separability between \gls{id} and \gls{ood} samples. 

\vspace{-0.3cm}
\subsection{Approaches based on Distance Functions}

The intuition behind distance-based \gls{ood} detection is that the latent representation of the \gls{id} and \gls{ood} samples can be divided into separate clusters. As such, the distance between representations of \gls{id} samples should be smaller than the distance between \gls{id} and \gls{ood} representations. The typical distances used are \textit{Mahalanobis distance}, \textit{Hamming distance}, and \textit{L\textsubscript{2}} distance. As these approaches typically use class-conditional prototypes, they are a form of \textit{prototype learning}. 

Lee \textit{et al.}~\cite{lee2018simple} proposed to model the features using class-conditional Gaussian distributions and use the Mahalanobis distance as the score for \gls{ood} detection. The distance is calculated for each layer separately, while auxiliary \gls{ood} data is used to fit the distances collected from different layers into a logistic regression model which greatly diminishes its utility as it requires access to the \gls{ood} distribution to be detected and trains a separate logistic regression model for each \gls{ood} distribution. In contrast, \cite{deepneighborsun2022} proposes to use L\textsubscript{2} distance of the test sample to the $k$-th nearest neighbor from the training dataset, thus leading to a DNN- and \gls{ood}-agnostic approach. The authors show that the performance of the KNN based \gls{ood} detector improves if the model is trained with contrastive learning. This is because the features form tighter clusters and the distance of the \gls{ood} data from any particular cluster increases.

Ming \textit{et al.} \cite{ming2022delving} proposed an approach based on cosine similarity, where they measure the alignment of the features of an input image with concept vectors obtained from the language model. The similarity score is then used in place of logit to calculate the softmax probability over the concepts from the language model. The similarity represents the probability of the image aligning with a specific concept. The maximum of these probabilities work as the score for \gls{ood} detection. The work in \cite{du2022siren, ming2023exploit} uses von Mises-Fisher distribution to model the class prototypes in hyper-spherical embedding space. The recent work \cite{bai2024hypo} used a similar loss function as \cite{ming2023exploit} and showed that it improves the \gls{ood} generalization performance of \gls{dnn}. Conversely, Lu \textit{et al.}~\cite{lu2023learning} showed that using a single class prototype is detrimental to learning the representation properly. The authors propose to use multiple prototypes for each class and assign each sample to a prototype using soft assignment. They also model the clusters using \gls{vmf} and use Mahalanobis distance as their \gls{ood} scoring metric. Gomes \textit{et al.}~\cite{gomes2022igeood} proposes to use geodesic distance between two data distributions -- i.e., the Fisher-Rao distance \cite{atkinson1981rao} -- arguing that this information geometric distance is more suitable for measuring difference between two distributions than the Mahalanobis distance.  Finally, Olber \textit{et al.}~\cite{olber2023detection} proposed an approach based on the concept of \textit{neural activation pattern}. Specifically, the authors extracted activations from all layers of a \gls{dnn}, binarize them based on a threshold value, and concatenate them in a vector. If the Hamming distance of this binary vector, which represents the activation pattern of the \gls{dnn}, to the nearest \gls{id} sample is greater than a certain threshold, the input sample is considered as an \gls{ood} sample.  \vspace{-0.5cm}

\subsection{Other Approaches}

The work in \cite{liu2023neuron} proposed \gls{nac}, which considers the difference in behavior of neurons for \gls{id} and \gls{ood} samples. Specifically, \gls{nac} measures the frequency of activation of a neuron for training data in the form of coverage scores. The intuition is that \gls{ood} samples trigger neurons that are not usually triggered by \gls{id} data. The coverage score  measures the alignment of the triggering of the neurons with training data and should be higher for \gls{id} than \gls{ood} input. Although  inspired by \cite{sun2021react,gradnorm2021}, the key innovation of this work is that \gls{nac} may also be used as a regularizer during training to improve \gls{ood} generalization. 

Yi \textit{et al.}~\cite{yi2021improved} connected \gls{ood} generalization performance to shift in distribution. It also showed that the generalization bound tightens for \gls{dnn} robust against perturbation to the input. The authors proposed to use adversarial training to make the \gls{dnn} robust and showed that the performance of \gls{dnn} improves when adversarially trained. This work established a connection between adversarial robustness and \gls{ood} robustness and provided the insight that the problem of resilience of \glspl{dnn} against intentional and unintentional perturbation is connected.

The work in \cite{Morningstar2020DensityOS} estimated the typicality of selected test-statistics $T_n(x)$ using \glspl{kde} on sets of statistics and creating a density estimate for individual samples by using product-of-experts (POE) structures. The POE provides the probability that an input is jointly typical for all test statistics. The \gls{id} samples are typical for all test statistics providing higher probability while \gls{ood} samples provide lower probability as they are not typical for all test statistics. The authors use different test statistics like posterior/prior cross-entropy, posterior entropy , posterior/prior KL divergence, posterior expected log-likelihood, and so on.

Finally, Huang \textit{et al.}~\cite{huang2022densityregularization} proposed two density regularization methods, i.e.,  density consistency regularization and contrastive distribution regularization, to reliably calibrate and estimate sample density to identify \gls{ood} samples. Density consistency regularization enforces the agreement between analytical and empirical densities of categorical labels, while contrastive distribution regularization helps separate the densities between \gls{id} and \gls{ood} samples. The logarithm of the sample density function is used as the score function for \gls{ood} detection. \vspace{-0.5cm}

\section{Why are Intentional and Unintentional Perturbation Connected?}

A fundamental similarity between intentional and unintentional perturbations is that their effect is similar. Indeed, both can be modeled as a perturbation $\mathbf{\delta}$ such that for a \gls{dnn} $f$ and input image $\mathbf{x}$, $f(\mathbf{x}+\mathbf{\delta}) \neq f(\mathbf{x})$. \textit{This suggests that intentional perturbation can be modeled as \gls{ood} examples}. This has been explored in \cite{deng2021libre}, where the adversarial inputs are modeled as a special case of \gls{ood} samples. The presence of natural adversarial samples \cite{hendrycks2021nae} strengthen the support for connection between intentionally and unintentionally perturbed samples.

Another similarity  is that the \gls{dnn} becomes overconfident in the case of both intentional and unintentional perturbation. This indicates that detection algorithms will share common characteristics. Prior work also reflects this intuition, as they attempt to present a common detection framework \cite{lee2018simple, raghuram2021general}. On the other hand, the two communities for detection of intentional and unintentional perturbation can also benefit from exploring \textit{cross-domain} approaches. For example, the detection of intentionally perturbed samples utilize the Bayesian Uncertainty estimation model \cite{feinman2017detecting, deng2021libre}, yet this approach has not been explored for \gls{ood} detection. Moreover, intentional perturbation detection use class-conditional Gaussian distribution \cite{lee2018simple}, mixture of Gaussian \cite{Zheng2018}, kernel density estimation \cite{feinman2017detecting}, and \gls{lid} \cite{ma2018characterizing}. The community for unintentional perturbation detection has explored additional distributions - \gls{vmf} \cite{mardia2009directional} for example which is better at modeling embedding on a hypersphere. Another example of \textit{cross-domain} opportunity is the modeling of the activation patterns for intentional and unintentional perturbation in terms of discrete states. The work in \cite{lu2017safetynet} models the activation from the penultimate layer using binary or quaternary codes,  while \cite{olber2023detection} extracts binary neural activation pattern from multiple layers to characterize the label shift. These connections signify that the detection of intentional perturbation can benefit from exploring approaches adopted in unintentional perturbation detection. On the other hand, the assumption of access to perturbed samples is a common bottleneck that can be jointly explored. This leads to incomplete modeling of the perturbed samples. 

Another connection lies in the intermediate representation in \gls{dnn}. For example,\cite{bai2021improving} shows that adversarial inputs tend to activate all filters more or less uniformly. A similar phenomenon is observed in case of unintentional perturbation, especially in the case of label shift, where the \gls{ood} samples show unusually high activations for some filters. Such samples can be effectively detected by reshaping the filters activations through suppression \cite{sun2021react}, typical feature selection \cite{zhu2022boosting} and feature reshaping optimization \cite{xu2023vra}. Since mitigating overconfidence issue improves the detection of unintentional perturbation \cite{wei2022mitigating}, we need to explore such approach also for intentional perturbation detection. \vspace{-0.3cm}

\vspace{-0.1cm}
\section{Summary of Existing Research Challenges In Perturbation Detection}

\textbf{Security and Privacy Issues:}~The work in \cite{chen2022relaxloss} showed that a large difference between the training loss and the testing loss can lead to membership privacy risk, i.e., the adversary may increase their ability to conclude that
an entity is in the input dataset \cite{li2013membership}. Traditional training algorithms provide high confidence on both training and testing samples provided the \gls{dnn} has low generalization gap, which leads to overconfidence. In order to alleviate the overconfidence issue, some approaches propose modified training loss to facilitate the detection of adversarial or \gls{ood} examples \cite{chyou2023unsupervised, choi2023balanced, liu2020energy}. While these approaches reduce the test error, the change in generalization gap is neglected, which introduces possible  privacy vulnerabilities. For example, \cite{chyou2023unsupervised} imposes specific pattern for unperturbed samples and any deviation from the pattern is detected as adversarial sample. The effect on privacy is not studied as the generalization gap with this training approach is not mentioned. If this approach increases divergence between the training loss distribution and test loss distribution, then the privacy of the DNN would be affected. 

\noindent \textbf{Research in Domains Different from Multi-class Classification:}~The \gls{ood} detection literature is predominantly focusing on \glspl{dnn} for multi-class classification. Settings such as multi-label classification, object detection, segmentation tracking are largely understudied. We were able to find only the work \cite{hsu2020generalized, choi2023balanced} focusing on unintentional perturbation for segmentation tasks and only the work \cite{du2022vos, du2022siren, Wilson_2023_ICCV, du2022stud} focusing on unintentional perturbation object detection. 

\noindent \textbf{Detecting Perturbations at Scale:}~Most of the work in \gls{ood} detection reports detection performance on CIFAR10 and CIFAR100 benchmarks. On the other hand, the work in \cite{Hendrycks2022ScalingOD} has shown that \gls{ood} detection performance degrades significantly when considering real-world datasets. There is also a varying degree of performance as a function of the \gls{ood} dataset and architecture. For example, \cite{liu2023gen} reports the \gls{fpr} for Imagenet-1k benchmark varying from 22.60\% to 54\% in the Swin transformer architecture, while in the BiT-S-R101$\times$1 architecture it varies from 80.35\% to 97.25\%. This variability hints at a current lack of  a dataset-agnostic \gls{ood}  detection approach. Another key issue is determining which detector might work for a specific task during deployment. The average performance of the detectors would be unreliable in real-world as one would not know the distribution of \gls{ood} samples and might encounter worst-case distributions on which the detector provides high \gls{fpr}. This opens up another research question: can we create a detector performance whose performance is \gls{ood} distribution agnostic?   

\noindent \textbf{Computation vs Performance Trade-off:}~The aspect of application and hardware specific constraints (e.g. computational cost, latency, energy etc.) is mostly ignored in the current literature of perturbation and semantic shift detection. There is no benchmark for latency, and computation cost metrics and these are rarely reported. \cite{picot2023a} is the only work we could find that reports the computation time for detection of intentional perturbation but there is no analysis of the computational cost or energy expenditure. These are important metrics for implementation in edge devices. As a result, it is important that these metrics are reported as well. 

\noindent \textbf{Lack of Curated Dataset for OOD Detection:}~Most of the OOD detection literature focuses on semantic \gls{ood} and on the generalization capability of the \gls{dnn} for non-semantic \gls{ood}. An interesting observation was recently reported in\cite{yang2024imagenetood}, where the authors introduced the Imagenet-OOD dataset to separatesemantic and covariate shifts. Moreover, the authors showed that existing methods are susceptible to detecting covariate shift rather than semantic shift. This suggests that existing methods are mostly designed for semantic \gls{ood}. This calls for curated datasets that (i) disentangle covariate shift and semantic shift and (ii) are sufficiently large-scale to capture real-world scenarios. 

\noindent \textbf{Lack of Explainability:}~The existing literature lacks an explainable way to separate semantic and non-semantic \gls{ood}. Specifically, current approaches to\gls{ood} detection cannot explain which features of the input lead the \gls{dnn} to generate the  embeddings or the scores used to perform the detection. A more explainable \gls{ood} detection method can provide insights and drive further innovation.

\noindent \textbf{Joint Resilience to Intentional and Unintentional Perturbation:}~In our view, guaranteeing resilience to intentional and unintentional perturbations cannot be seen as separate issues. Due to the disjoint communities of adversarial robustness and \gls{ood} robustness, the problem of joint resilience against both types of perturbations has been under-explored. The only work we have found is \cite{zou2023on}. Although the work shows impressive performance on five different datasets, the number of images in different datasets and the variety is not sufficient to represent real-world conditions. The PACS \cite{li2017deeper}, VLCS \cite{fang2013unbiased}, and OfficeHome \cite{dubois2021optimal} datasets used contain 7, 5, and 65 classes with 9991, 10729, and 15588 example images respectively. To experiment on more realistic settings, a reasonable approach would be to consider datasets like DomainNet \cite{peng2019moment} and ImageNet-C \cite{hendrycks2019robustness}.      

\section{Conclusions and Call to Action}

Over the last few years, a substantial amount of research has  identified various issues regarding the resilience of \glspl{dnn}, including susceptibility to adversarial attacks and distributional shifts. This  has spurred a significant amount of research aimed at enhancing the robustness and reliability of \glspl{dnn}. Efforts to bolster resilience have led to the development of novel techniques such as adversarial training, robust optimization, and domain adaptation. These approaches aim at making DNNs more robust against potential threats, while also improving their generalization across operational settings. The ongoing research into the underlying mechanisms of vulnerability can only advance through a collaborative effort from both intentional and unintentional perturbation detection communities. In this work, we have surveyed the state of the art in both intentional and unintentional perturbation detection, with the key aim of consolidating the approaches being employed. We have also listed a set of challenging research directions in this field. We hope that this survey will spur excitement in both communities and set new research directions in this topic.  

\vspace{-0.3cm}
\bibliographystyle{ACM-Reference-Format.bst}
\bibliography{bib-francesco,bibliography,attack,domain_adapt,advtrain, reference}

\end{document}